# Light-induced insulator-metal transition in $Sr_2IrO_4$ reveals the nature of the insulating ground state

Dongsung Choi[1†], Changming Yue[2†], Doron Azoury[3], Zachary Porter[4,5], Jiyu Chen[2], Francesco Petocchi[2], Edoardo Baldini[3,6], Baiqing Lv[3,7], Masataka Mogi[3], Yifan Su[3], Stephen D. Wilson[4], Martin Eckstein[8], Philipp Werner[2*], Nuh Gedik[3*]

[1]Department of Electrical Engineering and Computer Science, Massachusetts Institute of Technology, Cambridge, MA, USA.

[2]Department of Physics, University of Fribourg, 1700 Fribourg, Switzerland.

[3]Department of Physics, Massachusetts Institute of Technology, Cambridge, MA, USA.

[4]Materials Department, University of California Santa Barbara, Santa Barbara, CA, USA.

[5]SLAC National Accelerator Laboratory, Stanford University, Stanford, CA, USA.

[6]Department of Physics, The University of Texas at Austin, Austin, TX, USA.

[7]School of Physics and Astronomy, Shanghai Jiao Tong University, Shanghai, People's Republic of China.

[8]Department of Physics, University of Erlangen-Nürnberg, 91058 Erlangen, Germany.

[†]These authors contributed equally: Dongsung Choi, Changming Yue

*e-mail: gedik@mit.edu, philipp.werner@unifr.ch

**Abstract**

$Sr_2IrO_4$ has attracted a lot of attention due to its structural and electronic similarities to $La_2CuO_4$ which is the parent compound of high-$T_c$ superconducting cuprates. It was proposed to be a strong spin-orbit coupled $J_{eff} = 1/2$ Mott insulator, but the Mott nature of its insulating ground state and the origin of the gap have not been conclusively established. Here, we use ultrafast laser pulses to realize an insulator-metal transition in $Sr_2IrO_4$ and probe the resulting dynamics using time- and angle-resolved photoemission spectroscopy. We observe a closing of the gap and the formation of weakly-renormalized electronic bands in the gap region.

Comparing these observations to the expected temperature and doping evolution of Mott gaps and Hubbard bands provides clear evidence that the insulating state does not originate from Mott correlations. We instead propose a correlated band insulator picture, where antiferromagnetic correlations play a key role in the opening of the gap. More broadly, our results demonstrate that energy-momentum resolved nonequilibrium dynamics can be used to clarify the nature of equilibrium states in correlated materials.

**Introduction**

Proposed as a $J_{eff} = 1/2$ Mott insulator (*1*) resulting from strong spin-orbit coupling (SOC), $Sr_2IrO_4$ has been considered as an alternative material platform to explore the mechanism of high-$T_c$ superconductivity, due to its structural and electronic similarities to $La_2CuO_4$ which is the parent compound of high-$T_c$ superconducting cuprates (*2*). While the $J_{eff} = 1/2$ ground state (*3*) and the important role of SOC in the opening of the charge gap (*4*) in $Sr_2IrO_4$ are experimentally confirmed, the Mott nature of the low-temperature insulating phase has not been firmly established. In fact, there are multiple experimental observations that are inconsistent with the Mott picture. For example, as temperature increases, the optical gap decreases anomalously fast compared to other semiconductors and closes at around 600 K (*5*). The strong temperature dependence of the optical gap indicates that the gap closing is not simply a thermal effect (*5*). Also, 600 K is too low for the thermal closing of the Mott gap (the gap at the M point for $Sr_2IrO_4$ is ~ 0.32 eV). As an alternative to the Mott picture, the Slater mechanism, which opens the gap via a long-range antiferromagnetic (AFM) order, has been proposed (*6–8*), but this also fails to explain some of the observations (*5, 9–11*). For instance, the temperature dependent resistivity (or resistance) shows insulating behavior at least up to 300 K – 350 K (Néel temperature ($T_N$) ~ 240 K) (*9, 10*), which means that the gap is still open above $T_N$. The optical gap also persists above $T_N$ (*5*) and a non-zero gap was observed via angle-resolved photoemission spectroscopy (ARPES) at room temperature (*11*). Therefore, neither the Mott nor the Slater mechanism is fully consistent with the experimental observations and the nature of the gap in $Sr_2IrO_4$ is still highly controversial.

In this work, we used time-resolved ARPES (trARPES) to investigate the evolution of the gap and the electronic bands as a function of electronic temperature and photo-doping via light excitation. We observed that the gap is closed and weakly-renormalized electronic bands are formed in the gap region as a result of high electronic temperature and photo-doping. We show that these experimental results, in combination with many-body calculations, reveal the origin of the insulating ground state.

**Results**

Single crystal $Sr_2IrO_4$ samples were grown through the flux method (*12*). The crystal structure is shown in Fig. 1A. Known to be isostructural to $La_2CuO_4$, a parent compound of high-$T_c$ superconducting cuprates, $Sr_2IrO_4$ has a $K_2NiF_4$ structure, a layered perovskite. The unit cell is a $\sqrt{2} \times \sqrt{2} \times 2$ supercell of the in-plane square lattice due to the distortion in the Ir-O-Ir bonding angle (*13*). This leads to a folding of the Brillouin zone as shown in Fig. 1B. $Sr_2IrO_4$ exhibits an insulating AFM state below $T_N \sim 240$ K.

Time-resolved ARPES (trARPES) measurements on $Sr_2IrO_4$ were performed with a time-of-flight type detector which allows a simultaneous recording of the 3D band structure as a function of energy (E) and two in-plane momenta ($k_x$ and $k_y$). The samples were cleaved in the ARPES chamber at a pressure of mid $10^{-11}$ torr at ~40 K. We excited $Sr_2IrO_4$ with 1.2 eV light pulses and recorded the subsequent evolution of the electronic band structure using 21.6 eV probe pulses with an energy resolution of ~36 meV. The measurement window in momentum space is about 1 Å$^{-1}$ in diameter. trARPES has unique advantages compared to static ARPES in that it can probe the electronic band structure below and above the Fermi level at high electronic temperature while the lattice is still cold (*14*) (see section SE8 in the experimental part of Supplementary Information).

We first characterized the sample with static ARPES, as shown in Figs. 1C and E. After parking our measurement window at the X point, in Fig. 1C, the four bands connecting the X point and the surrounding M points show the characteristic spectral features of $Sr_2IrO_4$ (*15–17*). Fig. 1D and F show that the calculated spectra for the $J_{eff} = ½$ manifold obtained by the two-site oriented cluster dynamical mean field theory (OC-DMFT) (*18*), which captures short-range antiferromagnetic correlations, well reproduce the experimental spectral features in Fig. 1C and E. Note that the simulation does not include $J_{eff} = 3/2$ states. In SE1, an energy distribution curve (EDC) is taken at the X point and we compared the fitting results to those from our simulation and the literature.

The temporal evolution of the electronic band structure after the perturbation by the pump pulse is displayed in Fig. 2A. At the delay time corresponding to the maximum pump-probe signal (PPm), new spectral weight with clear band features is generated in the gap. At these energies, there is no occupation before the pump pulse arrives, as shown in PPm – 3 ps. Also, from Fig. 1E (as well as Fig. SE1B and C), it is confirmed that the static band structure is insulating with the valence band top located at ~ -0.11 eV, which means there is a gap before the pump pulses arrive, as expected. We attribute the faint spectral feature in Fig. 2A at $E_F$ at PPm – 3 ps to the surface photo-voltage effect and thermal broadening of the valence band due to the steady-state heating by the pump pulses (see SE9). At 1 ps after PPm, the gap is recovering. This observation is further confirmed by the EDCs at the branch and the X point, which show an occupation of the gap region by spectral intensity at PPm, followed by a relaxation, as illustrated in Fig. 2B and C. Note that the EDCs are normalized by the total electron counts of each snapshot. We do not observe any indication of a conduction band edge at energies $E > E_F$ in the EDCs at PPm. Such bands should manifest themselves as a hump feature (*19*) at $E - E_F \cong 0.1 - 0.2$ eV at the branch and $0.15 - 0.3$ eV at the X point from our simulation (see SE1).

To analyze the detailed shape of the electronic band structure appearing in the gap at PPm, we performed a Gaussian fitting on the momentum distribution curves (MDCs) at each energy in energy-momentum (E-k) cuts. The extracted FWHM is overlaid on the E-k cuts in Fig. 3B and C. This procedure helps to visualize the detailed structure of the excited states. In Fig. 3C, at PPm, it is re-confirmed that the excited states at the branch occupy the gap region by crossing $E_F$ and extending to a high energy over 0.1 – 0.2 eV where the conduction band is expected to be located in the unperturbed state. This is a clear and dramatic change compared to the situation at PPm – 3 ps in Fig. 3B. The shape of the excited state preserves a well-defined FWHM to high energies and yields a pillar shape. In Fig. 3E, the pillar shape from the experiment is reproduced by the calculated spectra from OC-DMFT simulations with electronic temperature and photo-doping level of 800 K and 8% electrons/Ir-site, respectively, which are close to the values (818 K and 7.9 %) extracted from the EDC at PPm and at the X point (see SE3 and SE4). A similar analysis was performed at the X point and the calculated spectra well explain the excited states in the experimental results, which is presented in Fig. SE5.

We note that our OC-DMFT simulation uses a single value for the on-site Coulomb potential $U_{eff}$. In previous studies, very different values of $U_{eff}$ were considered to describe the insulating state (1.1 eV (*11*) and 2 eV (*16*)) and the metallic state (0.6 eV (*11*) and 0 eV (*16*)). While some modifications in $U_{eff}$ can be expected due to the different screening environments, such a large change is not realistic. In section ST2.A in the theory part of Supplementary Information, we present the self-consistently computed effective interactions from GW+EDMFT calculations (*20*), which take screening effects into account, and which suggest that the temperature and doping induced modifications in $U_{eff}$ are at most 30%. In our model, we therefore fix the value of $U_{eff}$ to 1.5 eV. With the use of a numerically exact impurity solver in the OC-DMFT calculations (*21*), this single parameter allows us to reproduce the experimental results qualitatively, and even semi-quantitatively.

The above analysis shows two key features: (1) the non-trivial evolution of the dispersion of the excited states at PPm (the pillar shape at the branch in Fig. 3C and the biconcave shape at the X point in Fig. SE5C) and (2) the absence of the conduction band onset in Fig. 2B and C. These features demonstrate that the occupation of the gap region by the excited states at PPm is not consistent with a trivial gap-filling by the thermal broadening of the band. Even though the evolution of electronic spectrum below $E_F$ after photo-excitation is more complex (persistent hump at ~ -0.10 eV and slight shift of the peak at -0.24 eV to more negative energies) as shown in Fig. SE11, our observations on the excited states as a whole clearly indicate gap-closing and a transient light-induced insulator-metal transition at PPm (for a more detailed discussion, see SE11). This behavior, the gap-closing at PPm with clear features of the electronic band structure appearing in the gap region (Fig. 2 and 3C), is not expected for a typical Mott insulator. To illustrate the differences between $Sr_2IrO_4$ and a typical Mott insulator, we calculate the doping- and temperature-dependent spectral functions of a Mott insulator, considering the half-filled single-band Hubbard model on the square lattice at strong coupling. The single-site DMFT solution of this simple model allows us to reveal the essential physics, as shown in Fig. 4A - D. In a doped Mott insulator at low temperature, the system becomes a Fermi liquid metal with a sharp quasi-particle peak near the edge of the upper or lower Hubbard band, while the Mott gap persists, as shown in Fig. 4C. Similarly, the emergence of both hole- and electron-like quasi-particle states at the Hubbard band edges, with no coherent band features in the gap region, has been predicted in quasi-equilibrium states after photo-excitation (*22*). At high temperature with a certain level of electron doping, the system becomes a bad incoherent metal with a pseudo-gapped spectral function and partial filling-in of the Mott gap, while the quasi-particle peak disappears due to a very short life-time, as shown in Fig. 4D. A clear difference between $Sr_2IrO_4$ and a Mott insulator in a doped high-temperature state is that, in the experimental results and in the realistic simulations of $Sr_2IrO_4$, the gap is closed (Fig. 3C and E), while in a Mott insulator, the Mott gap is not closed but is partially filled in due to the broadening of the Hubbard bands (Fig. 4D). In particular, as temperature is increased in the Mott regime, no well-defined electronic bands re-emerge in the gap-region, contrary to what is observed in the constant energy cuts at PPm (Fig. 2A). This is a clear evidence for the non-Mott nature of the gap in $Sr_2IrO_4$. The closing of the gap together with the appearance of clear band features in the gap region (Fig. 2 and 3C), are reminiscent of the results for the weakly correlated system in Figs. 4E – H, calculated on the opposite side of the Mott transition or crossover in the phase diagram in Fig. 4I. Here, the gap is opened at low temperatures by antiferromagnetic correlations, while a metallic system with well-defined band structure is recovered upon doping or increasing temperature. In fact, the band features appearing in the gap region at PPm look similar to a weakly-correlated electronic band structure as obtained from density functional theory. The comparison to the model calculations in Fig. 4 thus suggests a non-Mott nature of the gap, but an important role of AFM correlations.

This insight is further supported by our GW+EDMFT results for realistic parameters and the analysis of the spin-spin correlation function and the self-energy in realistic OC-DMFT simulations (our $Sr_2IrO_4$ model). The details of the analysis below are presented in ST1.C, ST1.G, and ST2.B. The calculation of the spin-spin correlation function shows that AFM correlations drop in the temperature range of 200 K – 400 K, which is the range where the gap closes in the OC-DMFT simulations (see ST1.G and Fig. ST11). This indicates that the gap is induced by (short-ranged) AFM correlations. One may wonder if the two-site cluster used in OC-DMFT overestimates these correlations. However, in ST2.B and Fig. ST15, we also provide results from GW+(single-site-)EDMFT simulations which show that spectra similar to those of OC-DMFT are obtained if AFM long-range order is considered. Hence, AFM correlations -- short-ranged or long-ranged -- are essential for opening the gap and for reproducing the temperature and doping evolution of the spectra. The AFM gap in a weakly correlated system is not controlled by $U_{eff}$, but by the antiferromagnetic exchange, $J \sim t^2/U_{eff}$. In such a system, the gap is closed with weakly renormalized electronic bands appearing in the gap region once AFM fluctuations are quenched by either doping or temperature as shown in Figs. 4F – H. Correlations, however, do play a role at low temperature and low doping, since the frequency-dependent real part of the self-energy enhances the band splitting (see ST1.C). Based on these insights, and the good agreement between the experimental results and the theoretical calculations, we conclude that $Sr_2IrO_4$, instead of being a Mott insulator, should be regarded as a correlated band insulator (CBI) whose gap is induced by AFM spin correlations.

**Discussion**

Indeed, in the literature, there are experimental results supporting the CBI description of $Sr_2IrO_4$. In ref. (*5*), the authors measured the temperature dependent size of the optical gap using optical spectroscopy, and we can infer that the gap is closed at ~ 600 K by extrapolating the trend. Our theoretical estimation of the gap-closing at 400 K in OC-DMFT with short-ranged AFM correlations is roughly consistent with this finding (see Fig. ST4). An important point is that the gap survives above $T_N$ (~240 K), which is also directly confirmed by ref. (*11*) and consistent with refs. (*9*, *10*). This can be explained within the CBI picture, since the gap can be opened by strong short-ranged AFM correlations even if long-range AFM order is quenched (see paramagnetic OC-DMFT results in Fig. ST4). Another noteworthy aspect is that the rate with which the gap size decreases as temperature increases is anomalously fast compared to other semiconductors that they considered and this cannot be explained by a simple thermal effect. In fact, if the gap were of Mott nature, ~ 3700 K is required to thermally excite electrons across a gap of ~0.32 eV (at the M point) and destroy the insulating state (see ST1.G), which is significantly higher than the inferred gap-closing temperature of ~ 600 K. On the other hand, within the CBI picture, the gap induced by AFM correlations closes once these correlations are quenched by temperature, which is controlled by $J_{ex}$. More specifically, the in-plane spin correlations survive above the order transition (*23*) and the corresponding energy scale of about 60 meV (*24*) well matches the inferred gap-closing temperature of ~ 600 K.

In conclusion, our tr-ARPES measurements on $Sr_2IrO_4$ revealed a gap closing, with features of weakly correlated electronic bands appearing in the gap region, as a result of a high electronic temperature and photo-doping at PPm. Since such a behavior is not expected in a typical Mott insulator, our results provide strong evidence that the gap in $Sr_2IrO_4$ is not of Mott nature. The OC-DMFT solution of our theoretical model for $Sr_2IrO_4$ can qualitatively reproduce the measured spectra in the high-temperature and doped regime. Our theoretical analysis of the spin-spin correlation function and the self-energy suggests the picture of a CBI, where the gap at low temperature is induced by AFM correlations. Pump-probe ARPES provides access to high-temperature electronic states and doping dependent changes in the electronic structure, which can also be explored with realistic many-body simulations based on DMFT. Our study demonstrates how these state-of-the-art experimental and computational tools can be combined to reveal the nature of the ground state of a complex material. $Sr_2IrO_4$ in the ground state is a correlated band insulator whose gap is induced by AFM order. This CBI picture not only explains our trARPES data, but also clarifies results in the existing literature. Furthermore, the ability to access high electronic temperatures in trARPES can be used to answer questions on the role of AFM correlations in the opening of a gap or pseudo-gap in other materials such as cuprates or nickelates.

**Materials and Methods**

**Material synthesis.** Single crystals of $Sr_2IrO_4$ were synthesized via a flux method. Dry powders of $SrCO_3$ (99.99%), $IrO_2$ (99.95%), and anhydrous $SrCl_2$ (99.5%, all from Thermo Scientific) were mixed in a 2:1:5.5 molar ratio. The mixture was placed in a 100 mL Pt crucible with a loosely fitting Pt lid within a box furnace in an air environment. The temperature was raised to 1640 K, held for 5 hours, then cooled to 1120 K at 6 K per hour, before turning off the furnace and allowing it to cool to room temperature. Single crystals were removed from the crucible by dissolving the excess flux in water. The crystal phase was checked by x-ray diffraction using a PANalytical Empyrean diffractometer to exclude impurities of $Sr_3Ir_2O_7$ and $SrIrO_3$.

**Experimental setup.** The beam line of our trARPES setup starts from an Yb-fiber laser (Tangerine from Amplitude) generating pulses of a center wavelength of 1030 nm (1.2 eV), a duration of 135 fs, a pulse energy of 250 μJ, and a repetition rate of 300 kHz. The fundamental beam is split into a probe and a pump branch. In the probe branch, the second harmonic beam is generated by injecting the fundamental beam to a β-BBO crystal and it is focused on Ar gas ejected by a gas-jet nozzle to generate 21.6 eV (9[th] harmonic of the second harmonic of 1.2 eV). Since the seed beam and its harmonics are co-propagating after they are generated on the gas-jet, we selected 21.6 eV through a XUV monochromator (McPherson Inc) where gratings are installed with the off-plane mounted configuration (*25*). The selected 21.6 eV beam is focused on the sample through a toroidal mirror. The polarization of the probe beam is linear P-polarization. The polarization of the pump beam is linear and at the middle of S- and P-polarization (45 deg rotated clockwise from P-polarization if we look at the polarization in the direction of seeing the sample). The analyzer is of the angle-resolved time-of-flight (ARTOF) type (ARTOF 10k from ScientaOmicron). We used the 26-7 lens mode which provides an acceptance angle of ±13° and an energy window of ±3.5% of the center energy. The setup is the same as the one presented in ref. (*25*), but the laser, the module for generating the second harmonic of the fundamental beam, and the module for high-harmonic generation (HHG) are changed. In this study, we used the gas-jet method for HHG. The estimated energy resolution is about 36 meV (see SE6) and the estimated upper bound of the time resolution is about 400 fs (see SE7).

**Simulations with OC-DMFT and GW+EDMFT.** Details are described in the theory part of the Supplementary Information.

**Acknowledgements**

We are grateful to L. Balents, D. Hsieh, G. Refael, R. Averitt, S. Stemmer, A. Young, A. de la Torre, H. Ning, A. Zong, and N. Koirala for insightful discussions. DC acknowledges fellowship support from Ilju Academy and Culture Foundation. **Funding:** Work at MIT was supported by ARO MURI Grant No. W911NF-16-1-0361 (initial planning), Gordon and Betty Moore Foundation's EPiQS Initiative grant GBMF9459 (instrumentation) and US Department of Energy, BES DMSE (data acquisition and analysis). Theory and simulation work was supported by the Swiss National Science Foundation (Grant No. 200021-196966) and ERC Consolidator Grant No. 724103. SDW and ZP acknowledge support via NSF award DMR-1729489 (sample growth). **Author Contributions:** The project was conceived by NG. trARPES experiments were carried out by DC and DA. The experimental data was analyzed by DC. The OC-DMFT simulations were performed by CY and the GW+EDMFT simulations by JC and FP. The theory simulations were supervised by PW. The results were discussed by DC, CY, DA, JC, FP, ME, PW, and NG. The single crystal $Sr_2IrO_4$ samples were grown by ZP under the supervision from SDW. Discussions and trARPES maintenance were performed by DC, DA, EB, BL, MM, and YS. The manuscript was written by DC, CY, JC, PW, and NG. All the authors contributed to the final version of the paper. The entire project was supervised by NG. **Competing interests:** The authors declare that they have no competing interests. **Data and materials availability:** All data needed to evaluate the conclusions in the paper are present in the paper and/or the Supplementary Materials.13

# Figures and Tables

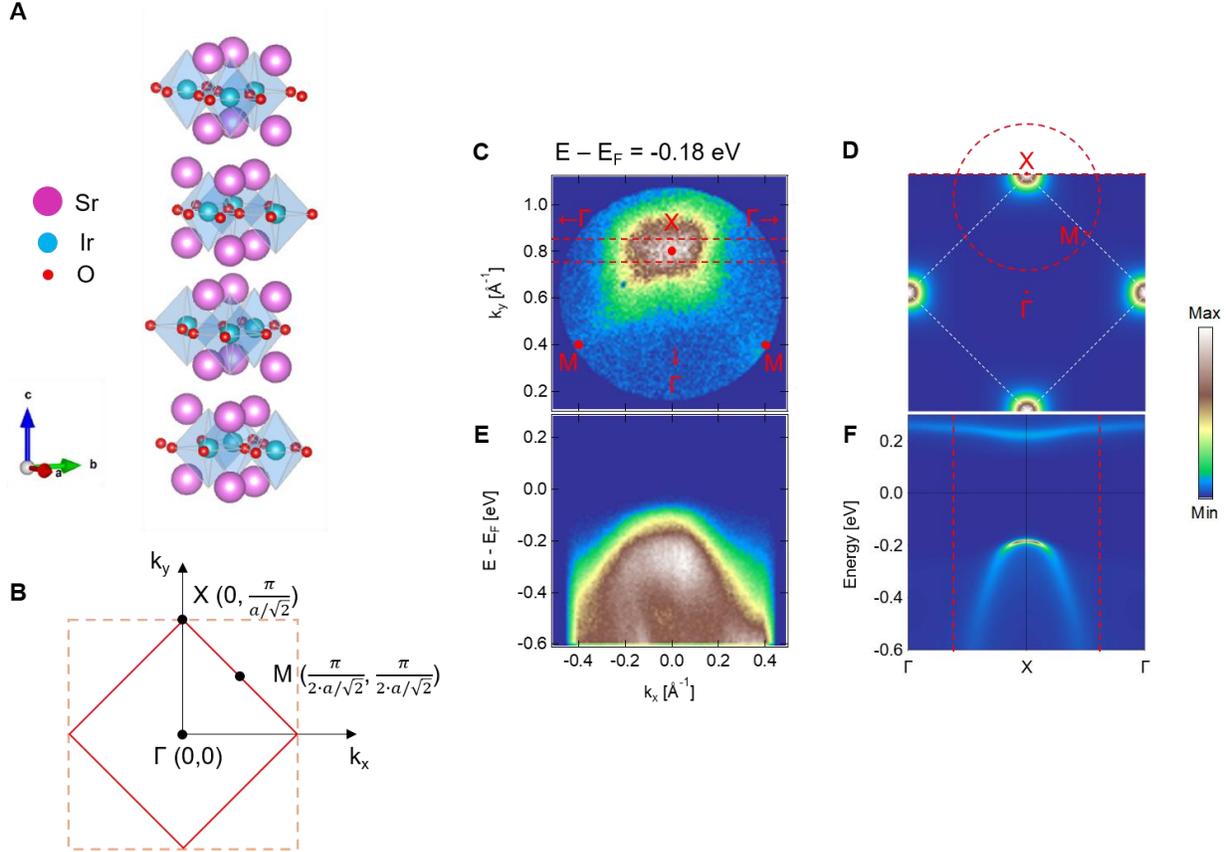

**Fig. 1. Static ARPES spectrum of $Sr_2IrO_4$.** (**A**) The crystal structure of $Sr_2IrO_4$. For the graphics, the visualization program VESTA was used and the source file for the crystal structure is from ref. (*26*). (**B**) The Brillouin zone of the electronic band structure of $Sr_2IrO_4$. The Brillouin zone is folded from the orange dashed square to the red solid diamond due to the distortion in the Ir-O-Ir bonding. Important high symmetry points, X, M, and Γ, are marked. The unit cell has the lattice constants of a ≅ 5.49 Å and c ≅ 25.78 Å for the in-plane and out-of-plane components, respectively (*26*). It is a $\sqrt{2} \times \sqrt{2} \times 2$ supercell of the in-plane square lattice with the lattice constant of $a_0 = a/\sqrt{2} \cong$ 3.88Å (*13*). (**C**) A constant energy cut at E – $E_F$ = -0.18 eV from a static ARPES spectrum of $Sr_2IrO_4$. The high symmetry points (the X and M points) and directions towards the Γ point are marked. (**D**) Calculated constant energy cut corresponding to (C). The red dotted circle shows the area corresponding to the measurement window in (C). (**E**) E-k cut along the Γ-X-Γ direction. The slicing location and the integration range are indicated by the pair of red dotted lines in (C). (**F**) Calculated E-k cut corresponding to (E) (spectral function, including the unoccupied band). Note that the simulation does not include $J_{eff}$ = 3/2 states. The red dotted lines indicate the momentum range corresponding to the measurement window in (E). The slicing location is indicated by the red dotted line in (D) (Γ-X-Γ direction). Note that the color scale in (C – F) is linear and ranges from the minimum value to the maximum value of the intensity in each panel. The color bar is on the right side of Fig. 1.

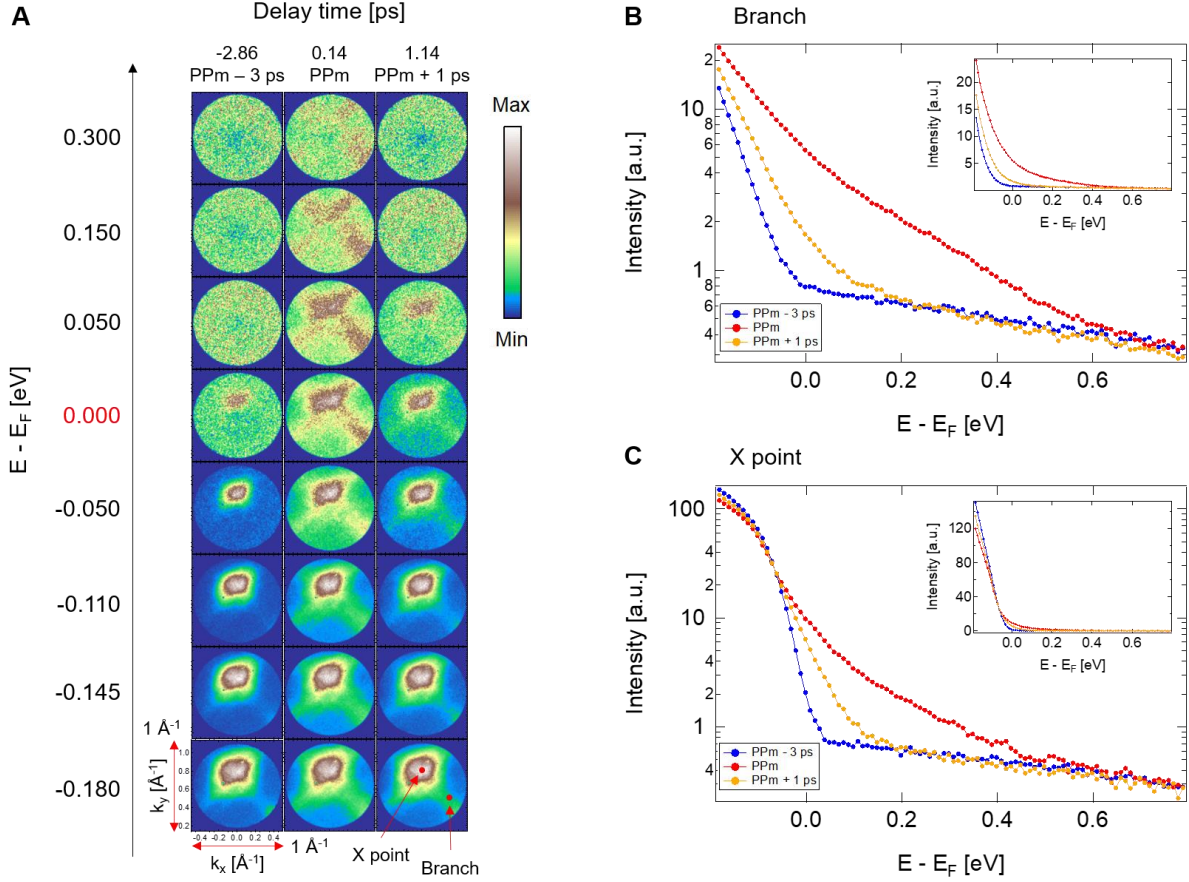

**Fig. 2. Light-induced spectral weight in the gap at PPm in $Sr_2IrO_4$.** (**A**) Constant energy cuts for different energies in pump-probe snapshots at different delay times. At PPm, new spectral weight with clear band features is generated in the gap above $E_F$ where originally there was no occupation before the pump pulse arrives (see PPm - 3 ps). At PPm + 1 ps, the gap is recovering. Note that at $E - E_F = 0$ eV and PPm – 3 ps, the spectral feature at the X point possibly originates from the surface photo-voltage effect and thermal broadening of the band due to the steady-state heating by the pump pulses (see SE9). (**B** and **C**) EDCs at the branch (B) and at the X point (C). The EDCs are normalized by the total electron counts of each snapshot. Since the total electron counts are on the order of $10^7$, we divided the total electron counts by $10^7$ and use these values for the normalization of each EDC. Note that the main panels are semi-log plots, while the linear-scale version is shown in the insets. The locations of the branch and the X point are shown in the constant energy cut at $E - E_F = -0.180$ eV of PPm + 1 ps in (A). The integration ranges of the EDCs are indicated in Fig. SE2. At the branch and the X point, the EDCs at PPm show spectral weight in the gap region and those at PPm + 1 ps show that spectral weight relaxes back, which re-confirms the observation in (A). In the EDCs at PPm, there is no indication of a conduction band or an upper Hubbard band, which would be expected to appear at $E – E_F \cong 0.1$ - 0.2 eV at the branch and 0.15 – 0.3 eV at the X point in the form of a hump feature, if it exists (see SE1).

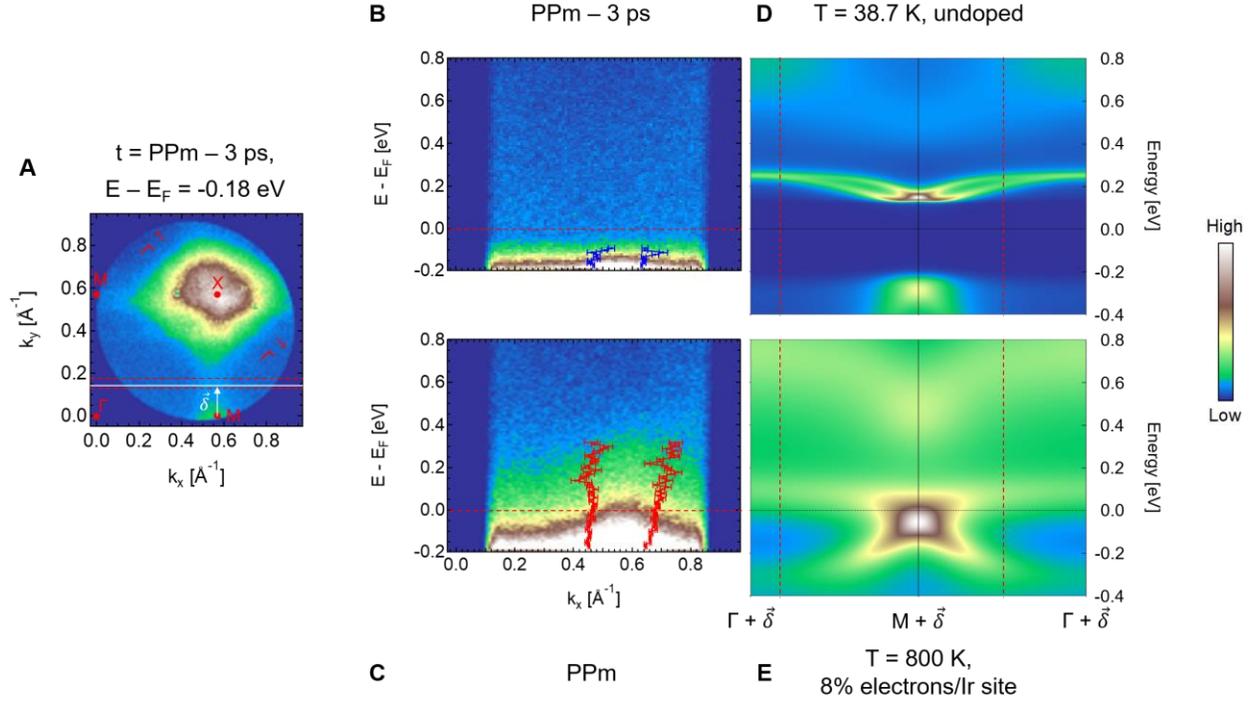

**Fig. 3. Features of the electronic band structure at PPm – 3 ps and PPm at the branch.** (**A**) A constant energy cut at PPm – 3 ps. Note that all the spectra in this paper were acquired with the orientation in (A). The spectra in Figs. 1C and 2A are rotated in the data analysis process for the convenience of generating E-k cuts along Γ-X-Γ and display purposes. (**B** and **C**) E-k cuts at PPm – 3 ps (B) and PPm (C). The slicing location and the integration range to generate E-k cuts are indicated by two red dotted lines in (A). Note that their color scale is saturated to clearly show the weak signals at high energies, but the range of the color scale is the same for both. The overlaid lines show FWHM (mean ± FWHM/2) of Gaussian fits on MDCs at each energy, which captures the detailed shape of the electronic band structure. At PPm, the excited states cross $E_F$ and extend to energies higher than ~ 0.1 – 0.2 eV where the conduction band is expected to be located in the un-perturbed state, which shows a clear change in the electronic band structure compared to that at PPm – 3 ps. Also, the shape of the excited state, which is pillar-like, preserves its FWHM to high energies. (**D** and **E**) Calculated E-k cuts at 38.7 K without doping (D) and 800 K with doping of 8% electrons / Ir-site (E). The slicing location is along the white line which is shifted by $\vec{\delta}$ (length = ¼ of X-M) from Γ-M-Γ in (A). The two red dotted lines in (D) and (E) indicate the momentum range corresponding to the signal window in (B) and (C). In (E), the gap is closed and the electronic band structure shows a pillar shape crossing over $E_F$, which reproduces the features at PPm in (C). Note that the ranges of the color scale in (A), (D), and (E) are set from the minimum to the maximum values of the intensity of each panel. The color scales in all the panels (A – E) represent a linear scale.

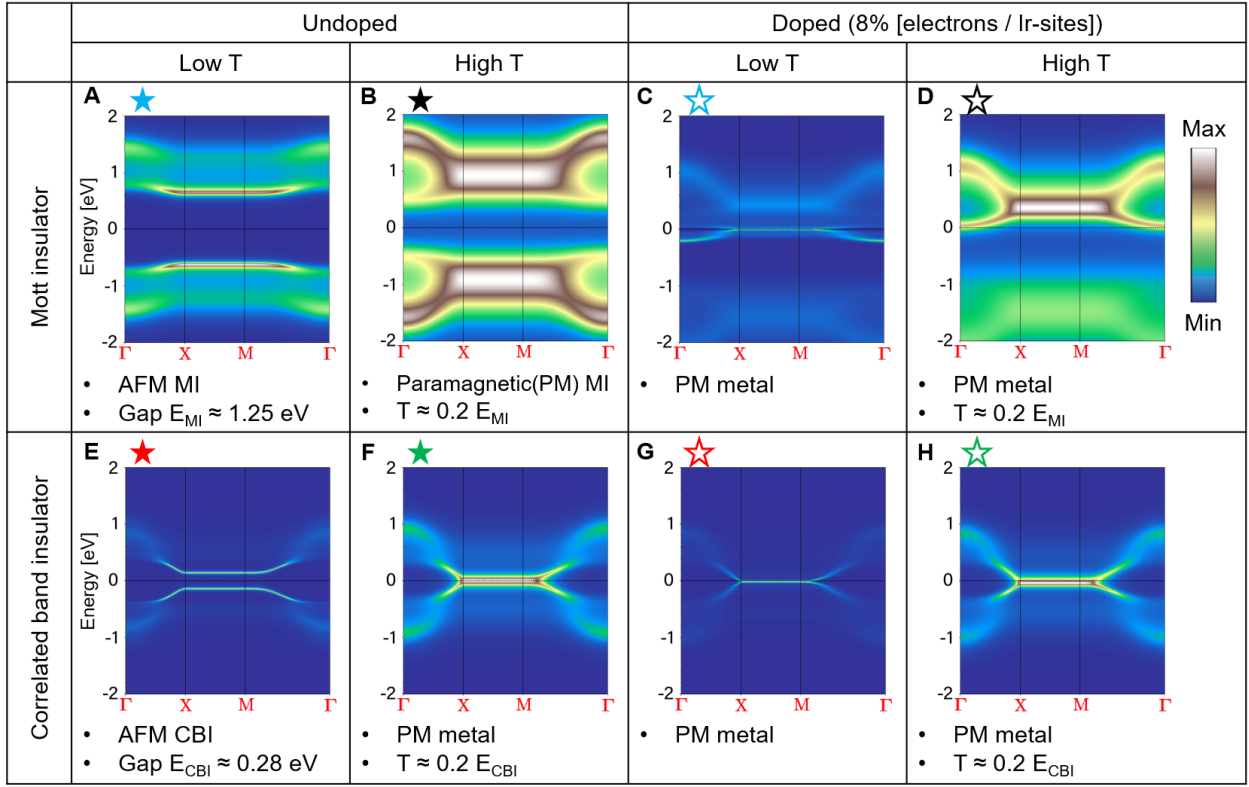

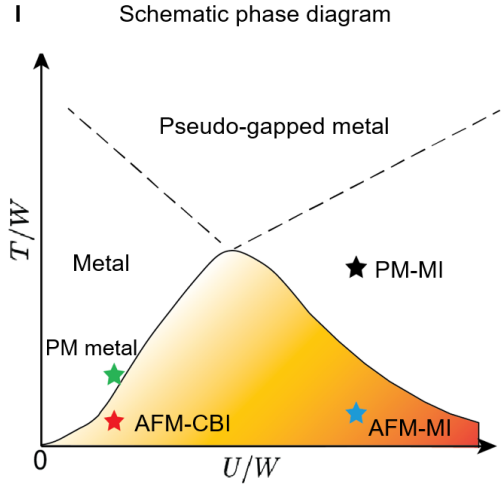

**Fig. 4. Comparison of undoped / doped and low / high temperature spectra for a Mott insulator and correlated band insulator.** Single-site DMFT results for a typical Mott insulator (MI) and a correlated band insulator (CBI), obtained for a half-filled single-band Hubbard model on the square lattice at strong and weak interaction U, respectively. Note that the model and the simulation method used here are simpler than in the realistic simulations of $Sr_2IrO_4$, but they suffice to explore the essential physics. At half-filling, the gaps at low temperature are $E_{MI} \approx 1.25$ eV and $E_{CBI} \approx 0.28$ eV for the MI and CBI, respectively. The momentum resolved spectra $A(\mathbf{k},\omega)$ in the first row (**A**, **B**, **C**, and **D**) represent the results for the MI (strong coupling regime, $U/W = 4/3$ where W is the bandwidth) and those in the second row (**E**, **F**, **G**, and **H**) show the results for the CBI (weak coupling regime, $U/W = 0.5$). Those in the first and second columns (A, B, E, and F), marked by the solid stars, correspond to the undoped systems and those in the third and the fourth columns (C, D, G, and H), marked by the empty stars, are for the doped systems (8% electrons / Ir site). Low temperature states in MI and CBI were simulated with the condition of $T \ll E_g$, where $E_g$ is the band-

gap. For the comparison between the high temperature states in MI and CBI, we choose T $\cong$ 0.2$E_g$ for both cases. (**I**) The schematic phase diagram for this model at half-filling in the space of T/W and U/W (adapted from ref. (*27*).). The yellow shaded region corresponds to AFM order, with the color gradient indicating a crossover from AFM-CBI to AFM-MI. Our results at PPm shown in Fig. 2 and 3C share essential features with (H) instead of (D)**,** where the system becomes a very bad incoherent metal with a pseudo-gapped spectral function and partial filling-in of the Mott gap, which shows the nature of the system to be a CBI rather than MI.

# Supplementary Information (theory part) for
# Light-induced insulator-metal transition in $Sr_2IrO_4$ reveals the nature of the insulating ground state

## ST1. TEMPERATURE- AND DOPING-DEPENDENT ELECTRONIC STRUCTURE OF SR$_2$IRO$_4$ FROM ORIENTED CLUSTER DYNAMICAL MEAN-FIELD THEORY

### A. Minimal two band Hubbard Model

We start from a realistic tight-binding (TB) model of Sr$_2$IrO$_4$ with spin-orbit coupling (SOC) [1]). There are six bands in the TB model involving $t_{2g}$ orbitals of two non-equivalent sites (denoted by $A$ and $B$) in the $\sqrt{2} \times \sqrt{2}$ unit cell. The corresponding band structure is shown in Fig. ST1. According to Ref. 1, if the Bloch wave functions of the TB model at each $\mathbf{k}$-point are projected on the basis of angular-momentum eigenstates $|j_{\text{eff}}, m_j\rangle$, one finds that the two low energy bands are primarily of $j_{\text{eff}} = 1/2$ character. The four bands far below the Fermi-energy are mainly formed by the $j_{\text{eff}} = 3/2$ states. The $|j_{\text{eff}} = 1/2, m_j\rangle$ states can be expressed as $|\frac{1}{2}, \frac{1}{2}\rangle_{A,B} = \frac{1}{\sqrt{3}} [|d_{xy\uparrow}\rangle + |d_{yz\downarrow}\rangle + i|d_{zx\downarrow}\rangle]_{A,B}$ and $|\frac{1}{2}, \frac{-1}{2}\rangle_{A,B} = -\frac{1}{\sqrt{3}} [|d_{xy\downarrow}\rangle - |d_{yz\uparrow}\rangle + i|d_{zx\uparrow}\rangle]_{A,B}$.

To obtain a minimal effective $j_{\text{eff}} = 1/2$ model, we have to project out the $|j_{\text{eff}} = 3/2\rangle$ states which are less relevant to the low-energy physics. This can be realized by performing the $N$th-order muffin-tin orbitals (NMTO [2]) band downfolding onto the $j_{\text{eff}} = 1/2$ basis. Here we choose $N = 1$ (first order). It turns out that the low-energy bands of dominant $j_{\text{eff}} = 1/2$ character are well reproduced, as shown by the red dashed lines in Fig. ST1. We note that the resulting basis functions (one basis per site) have a larger spatial extent than the original $j_{\text{eff}} = 1/2$ states, but are still Wannier-like and maximally localized [2]. Furthermore, they still have the specific $j_{\text{eff}} = 1/2$ characters. For simplicity, we still denote these downfolded basis states by $|j_{\text{eff}} = 1/2, m_j\rangle$.

There are two types of correlations treated in our theory. On the one hand, local correlations originating from the Coulomb repulsion are included by adding an on-site interaction $U_{\text{eff}}$ between electrons in the $|j_{\text{eff}} = 1/2, m_j\rangle$ orbitals. $5d$ electron systems exhibit more extended orbitals and more screening than $3d$ electron systems and the downfolding procedure leads to an additional spreading of the orbitals. For this reason, we use $U_{\text{eff}} = 1.5$ eV, which is smaller than typical values for more localized $3d$ orbitals. $U_{\text{eff}}$ is the only tunable parameter in our model and $U_{\text{eff}} = 1.5$ eV has been chosen such that all the temperature and doping dependent data provide a good match with available experimental results. In particular, we will not use different interaction parameters for undoped and doped compounds, as has been done in previous studies. For a discussion of the dynamical screening effect in self-consistent GW+EDMFT, we refer to Sec. ST2.

The full Hamiltonian can be written as

$$H = \sum_{\mathbf{k}\in\text{RBZ}} \sum_{m_j=-\frac{1}{2}}^{+\frac{1}{2}} \Psi^\dagger_{\mathbf{k},m_j} \begin{bmatrix} E_0 + t_1(\mathbf{k}) & \Delta_0 + t_2(\mathbf{k}) \\ \Delta_0 + t_2(\mathbf{k})^* & E_0 + t_1(\mathbf{k}) \end{bmatrix} \Psi_{\mathbf{k},m_j}$$
$$+ \sum_{i=1}^{N} \left[ U_{\text{eff}} n_{i,A,\frac{1}{2}} n_{i,A,-\frac{1}{2}} + U_{\text{eff}} n_{i,B,\frac{1}{2}} n_{i,B,-\frac{1}{2}} \right], \quad (1)$$

where $\Psi^\dagger_{\mathbf{k},m_j} = \begin{bmatrix} c^\dagger_{\mathbf{k},A,m_j}, & c^\dagger_{\mathbf{k},B,m_j} \end{bmatrix}$ and $i$ runs over all the unit cells. $t_1(\mathbf{k})$ (within the same sublattice) and $t_2(\mathbf{k})$ (between the $A$ and $B$ sublattice) are the hopping matrix elements determined numerically in the NMTO downfolding, which satisfy $\sum_{\mathbf{k}} t_{1,2}(\mathbf{k}) = 0$. The momentum $\mathbf{k}$ runs over the reduced Brillouin zone of the 2-sites unit cell. The onsite energy level $E_0 = 0.615$ eV and the hopping between the sublattices $\Delta_0 = -0.201$ eV are also determined in the downfolding.

The second type of correlations comes from the spatial degree of freedom. There is a strong inter-sublattice hopping $\Delta_0$. We will take into account short-ranged spatial correlations inside the unit-cell by treating a two-site cluster. For the sake of computational convenience, we transform the Hamiltonian (Eq. (1)) into the basis of bonding ($\alpha$) and antibonding ($\beta$) orbitals $|i, {}^\alpha_\beta, m_j\rangle = (|i, A, m_j\rangle \pm |i, B, m_j\rangle)/\sqrt{2}$. The kinetic part takes the form $\tilde{H}_0 = \sum_{\mathbf{k}\in\text{RBZ}} \sum_{\gamma=\alpha}^{\beta} \sum_{m_j=-\frac{1}{2}}^{+\frac{1}{2}} c^\dagger_{\mathbf{k},\gamma,m_j} \epsilon_{\gamma\gamma'}(\mathbf{k}) c_{\mathbf{k},\gamma',m_j}$ with $\epsilon_{{}^{\alpha\alpha}_{\beta\beta}}(\mathbf{k}) = t_1(\mathbf{k}) \pm \frac{t_2(\mathbf{k})+t_2(\mathbf{k})^*}{2}$ and $\epsilon_{\alpha\beta}(\mathbf{k}) = \frac{t_2(\mathbf{k})^* - t_2(\mathbf{k})}{2}$ for $\alpha \neq \beta$. In the bonding/anti-bonding basis, the local Hamiltonian becomes a two-orbital Hamiltonian with spin-flip

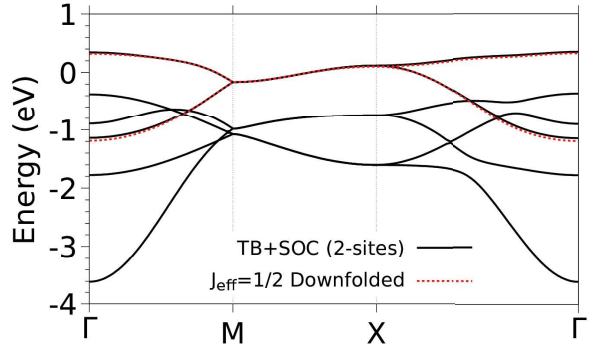

FIG. ST1. Band structure of the six-band TB model with SOC (black solid lines). The $NMTO$ downfolded $j_{\text{eff}=1/2}$ bands are shown by the red dashed lines, which well reproduce the low-energy bands near the Fermi energy. The lower four bands are mainly formed by the $j_{\text{eff}=3/2}$ orbitals which are integrated out in the $NMTO$ downfolding.

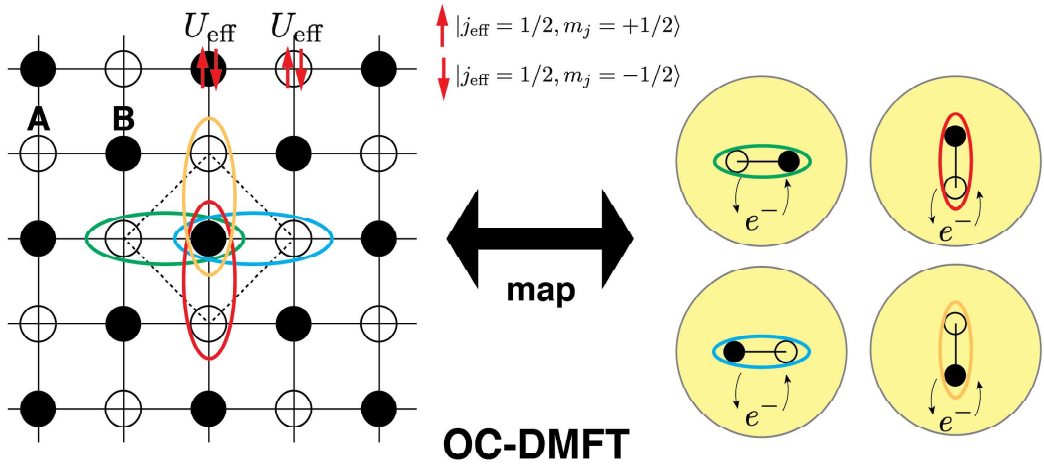

FIG. ST2. Schematic illustration of oriented cluster dynamical mean-field theory (OC-DMFT). The left part shows the Hubbard model on the square lattice with the $\sqrt{2} \times \sqrt{2}$ unit cell (indicated by the dashed box) containing two non-equivalent sites (empty circle for sublattice $A$ and solid circle for $B$). The hopping between sites can be long-ranged across several unit-cells. The on-site Coulomb interaction $U_{\text{eff}}$ acts between two electrons with oppsite pseudo-spin. $\uparrow$ represents the state $|j_{\text{eff}} = 1/2, m_j = +1/2\rangle$ and $\downarrow$ the state $|j_{\text{eff}} = 1/2, m_j = -1/2\rangle$. The green, red, blue and orange ellipses represent four choices of the cluster embedded in the lattice. In DMFT, the correlated lattice with a certian choice of cluster is mapped to a self-consistent impurity model with the cluster representing the 'impurity' and the yellow region representing the bath, as illustrated on the right hand side. In OC-DMFT, the results are averaged over the different choices of clusters.

and pair-hopping terms,

$$\tilde{H}^i_{\text{int}} = \sum_{m_j=-1/2}^{+1/2} \left[ (E_0 + \Delta_0) n_{i,\alpha,m_j} + (E_0 - \Delta_0) n_{i,\beta,m_j} \right]$$
$$+ U_c \sum_\alpha n_{i,\alpha\uparrow} n_{i,\alpha\downarrow} + U' \sum_{\alpha \neq \beta} n_{i,\alpha\uparrow} n_{i,\beta\downarrow}$$
$$- J_P \sum_{\alpha \neq \beta} c^\dagger_{i,\alpha\uparrow} c^\dagger_{i,\alpha\downarrow} c_{i,\beta\uparrow} c_{i,\beta\downarrow} - J_S \sum_{\alpha \neq \beta} c^\dagger_{i,\alpha\uparrow} c_{i,\alpha\downarrow} c^\dagger_{i,\beta\downarrow} c_{i,\beta\uparrow}, \quad (2)$$

with $U_c = U' = J_P = J_S = U/2$ [3–5].

## B. Oriented-cluster dynamical mean-field theory

The interacting lattice model $\tilde{H}_0 + \sum_i \tilde{H}^i_{\text{int}}$ is solved in the framework of dynamical mean-field theory (DMFT) [6] and its cluster extention [7], where the lattice is mapped to a two-orbital Anderson impurity model with a self-consistently determined electron bath and the local interaction as in Eq. (2). The numerically exact hybridization-expansion continuous-time quantum Monte-Carlo algorithm (CT-HYB) [8, 9] is adopted to treat the Anderson impurity model. In the bonding/anti-bonding basis, the hybridization matrix and self-energy turn out to be diagonal, which allows us to gain numerical efficiency and minimise the sign problem [10].

In the construction of model (1), we have to choose a two-site cluster. As shown by Fig. ST2, there are four possible choices (indicated by ellipses) for selecting an site of sublattice $A$ next to a given site of sublattice $B$. In cluster DMFT, the spatial correlations are only treated within the given cluster. After transforming the self-energy from the bonding and antibonding orbitals to the onsite $j_{\text{eff}}$ orbitals, besides the intra-site self-energies, there is only an inter-site self-energy between the $A$ and $B$ sites of the selected cluster. There is no inter-site self-energy perpendicular to the orientation of the selected cluster. As a result, a given choice of cluster breaks the 90° rotation symmetry. To restore the point-group symmetry, we adopt the oriented-cluster DMFT (OC-DMFT) method proposed in Ref. 13, which averages over the lattice Green's functions corresponding to the four differently oriented clusters indicated by the ellipses in Fig. ST2.

The non-local effects within the two-sites unit-cell are fully captured in our simulation using the numerically exact CT-HYB impurity solver. We note that the cluster impurity solver based on the Hubbard-I appoximation used in Ref. 13 overestimates the correlation effects compared with our solver for systems with the same $U_{\text{eff}}$. The authors of Ref. 13 had to choose different values for the insulator phase (1.1 eV) and for the metal phase (0.6 eV). However, in our simulations, we found that it is not necessary to adapt $U_{\text{eff}}$ to a given phase. While some doping or temperature effect on $U_{\text{eff}}$ can be expected (see Sec. ST2), it is likely smaller than 50%. We thus choose the same $U_{\text{eff}} = 1.5$ eV to reduce the number of adjustable parameters [14]. As shown by the following results, with this fixed interaction, our simulations can quantitatively reproduce the experimetal spectra, Fermi surfaces, and $T$-dependent gaps, not only in the insulator phase, but also for doped systems.

In the OC-DMFT calculations, we restrict the solution to the paramagnetic (PM) phase. As will be shown in Sec. ST2, at low-temperature, the momentum-resolved spectra in the PM phase are similar to those of a GW+(single-site)DMFT solution with anti-ferromagnetic (AFM) long-range order. In cluster-DMFT, calculations for the AFM phase are more expensive since the hybridization matrix in CT-HYB is no longer diagonal and the sign problem becomes severe. By analyzing the OC-DMFT results for the PM phase we can furthermore gain insights into the role of short-ranged AFM correlations.

## C. Temperature dependence of $A(\mathbf{k}, \omega)$ and gap closing in the undoped system

Figure ST3 shows the evolution of momentum-resolved spectra $A(\mathbf{k}, \omega)$ of the undoped system ($x = 0\%$) along the Γ-M-X-Γ **k**-path as the temperature $T$ is increased from $T =$ 38.7 K to $T =$ 1000 K. For a better comparison of the spectra and the bands, we have chosen a fixed color-bar ranging from 0 to 10 eV$^{-1}$ in all panels. The maximum value of $A(\mathbf{k}, \omega)$ is as large as 50 eV$^{-1}$ at $T = 38.7$ K and is as small as 5 eV$^{-1}$ at $T = 1000$ K. The bands shown in ST3**a** for $T = 38.7$ K are actually much sharper and narrower (see Fig. ST7) than it appears with the saturated color scale, which means there are well-defined quasi-particles emerging, albeit in the presence of a spectral gap. In a Mott insulator, sharp features such as spin-polaron bands can form in a low-temperature AFM system, but these are almost dispersionless. More importantly, as temperature is increased, we see the emergence of a broadened metallic band, with a dispersion very similar to that of the noninteracting system. This behavior is not expected in a Mott insulating system, where strong correlations lead to a pseudo-gapped state at high temperature. The observation of well-defined, dispersing quasi-particle bands with a separation much smaller than $U_{\text{eff}}$ at low temperature, and the

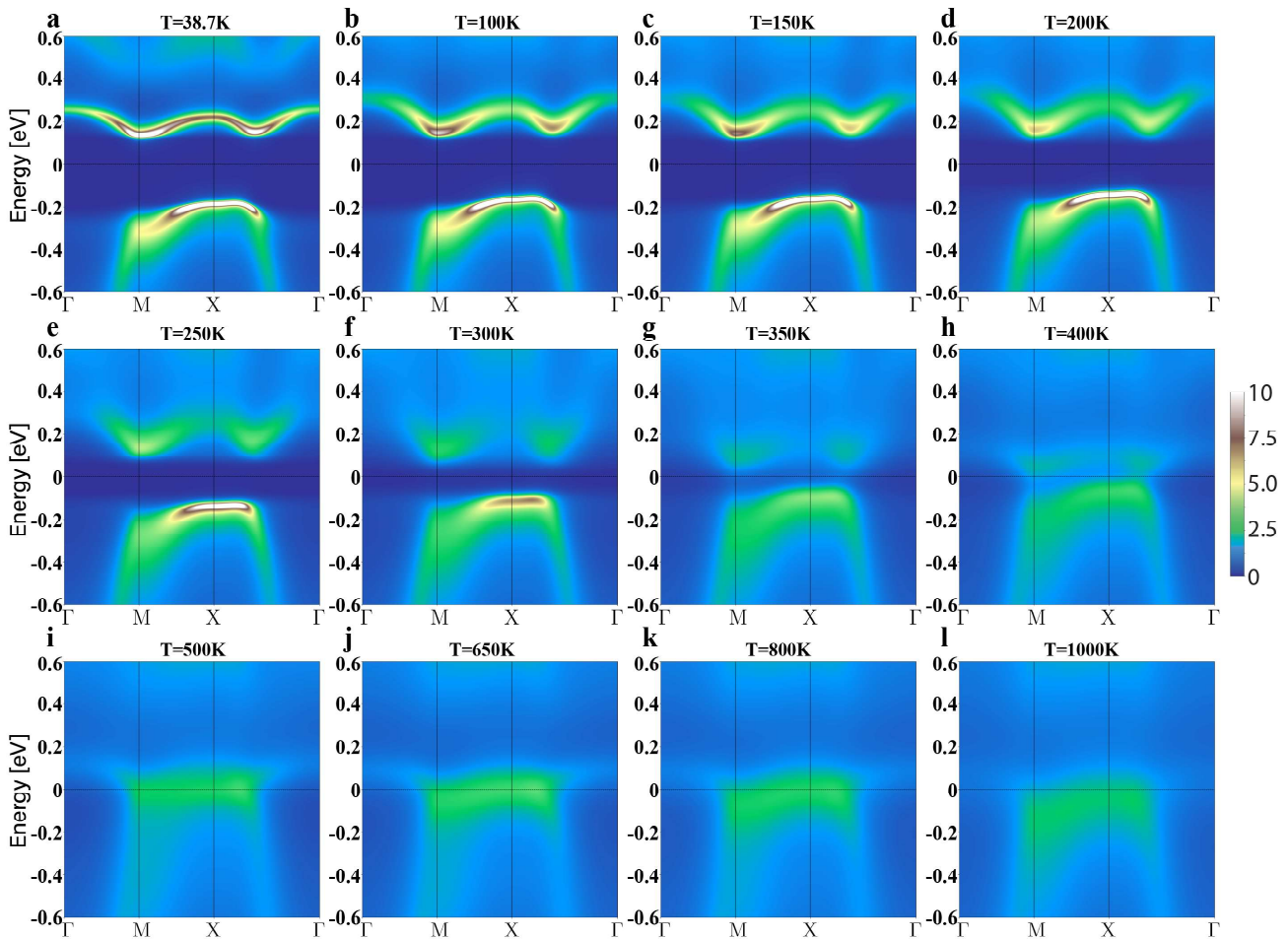

FIG. ST3. Momentum-resolved spetra along the Γ-X-Γ **k** path of the undoped system ($x = 0\%$) for $T = 38.7, 100.0, 150.0, \cdots, 1000$ K, respectively.

emergence of a weakly correlated metallic band at high temperature implies that the undoped system is not a Mott insulator, but rather a correlated band insulator.

We determine the gap size from the **k**-integrated spectra $A(\omega)$ (Fig. ST4**a,b**) using the method illustrated in Fig. ST4**c**. The $T$-dependent gap size obtained from the OC-DMFT simulations is shown by the green solid circles in Fig. ST4**d**. As one increases $T$, the gap size decreases slowly for $T \leq 150$ K then more rapidly above 150 K, and completely closes at $T = 400$ K, which is in semi-quantitative agreement with the optical gap determined by optical conductivity experiments [11] (note that the gap size at the lowest $T$ is $\sim 20\%$ smaller than the experimental value).

As shown by Fig. ST3**a-h**, as well as Fig. ST4**a**, the top of the valence band has much larger spectral weight than the bottom of the conduction band. The spectral weight of the valence-band top (conduction-band bottom) is mainly localized near the X-point (M-point) and is mainly originating from the bonding (anti-bonding) orbitals, as demonstrated by the orbital-projected $A(\mathbf{k}, \omega)$ (see Fig. ST5).

As temperature is increased beyond $T = 200$ K, the bonding band (valence-band top) quickly shifts upward and the anti-bonding band downward, which results in a closing of the gap around $T = 400$ K (see Fig. ST3**d-h**). This can be related to the temperature dependence of the real part of the self-energy. More specifically, the opposite shift of the two bands correlates with the temperature-dependence of the effective level splitting between the bonding and anti-bonding orbitals, which is enhanced by the real part of the self-energy $\Sigma$. In the non-interacting system with $\text{Re}\Sigma(\omega) = 0$ the band structure has no gap (see Fig. ST1). As one turns on interactions, the self-energy becomes nonzero and the real part develops a significant low-frequency dependence. This leads to a correlation enhanced splitting between the orbitals, roughly proportional to $\mathrm{Re}\Sigma_\beta(0) - \mathrm{Re}\Sigma_\alpha(0)$. The temperature dependence of this quantity is plotted in Fig. ST6a. As $T$ increases beyond 200 K, the correlation-induced splitting gets strongly reduced, which can be related to the fast decrease in the magnitude of the gap in Fig. ST3 and also in the **k**-integrated spectra shown in Fig. ST4b. Note that these results are for a PM system with only short-range AFM correlations; in a calculation with long-range AFM order, the real part of the self-energy would produce a spin-dependent shift.

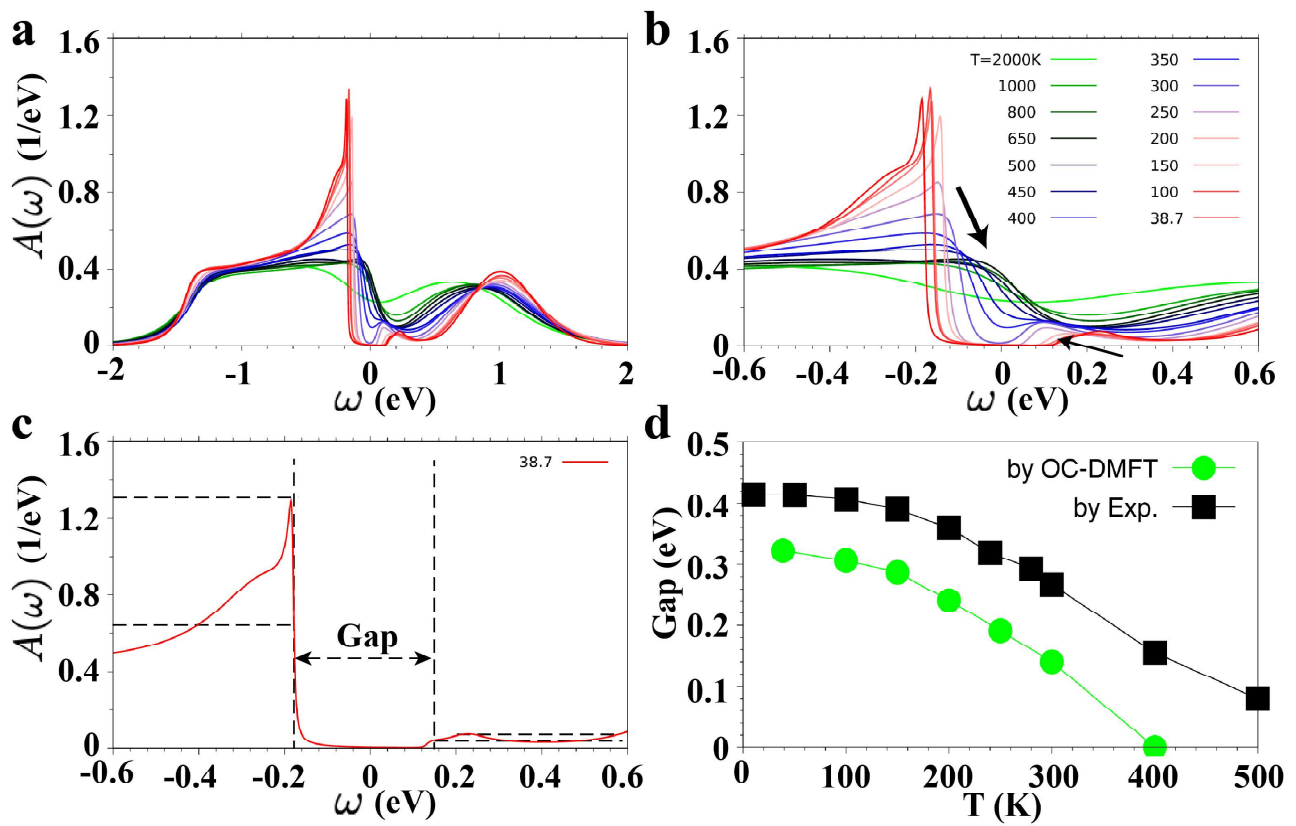

FIG. ST4. **a** (**b**) Momentum-integrated spetral functions of the undoped system for indicated temperatures in the energy window $-2$ to $+2$ eV ($-0.6$ to $+0.6$ eV). The thick (thin) arrow in **b** indicates that the quasi-particle peak below (above) the Fermi energy shift towards the Fermi energy. **c** Illustration of the method used to estimate the spectral gap. The half-maximum position of the low-energy peaks gives a good estimate of the spectral gap. **d** The estimated spectral gap as a function of $T$. The data point for $T = 350$ K is missing since there is no well defined gap in the spectral function (see panel **b**). The black squares in panel **d** plot the gap size determined from optical conductivity experiments (reproduced from Ref. [11]).

### D. Doping dependence of $A(\mathbf{k}, \omega)$ at low-temperature

In this subsection, we discuss the effect of electron doping on the electronic structure. The doping concentration $x\%$ corresponds to $x\%$ additional electrons per Ir atom.

Fig. ST7 shows the momentum-resolved spectra $A(\mathbf{k}, \omega)$ for $T = 38.7$ K along the $\Gamma$-M-X-$\Gamma$ **k**-path at doping concentrations $x = 0\%, 2\%, \cdots, 8\%$. Upon doping, the system becomes metallic with the anti-bonding band quickly shifted to the Fermi energy, resulting in an electron pocket near the M-point. The rapid shift of the bonding and anti-bonding band is again related to the doping dependence of $\mathrm{Re}\Sigma_\beta(0) - \mathrm{Re}\Sigma_\alpha(0)$, which is plotted in Fig. ST6b. This shift cannot be explained by the change in short-range AFM correlations, which are only slightly reduced between

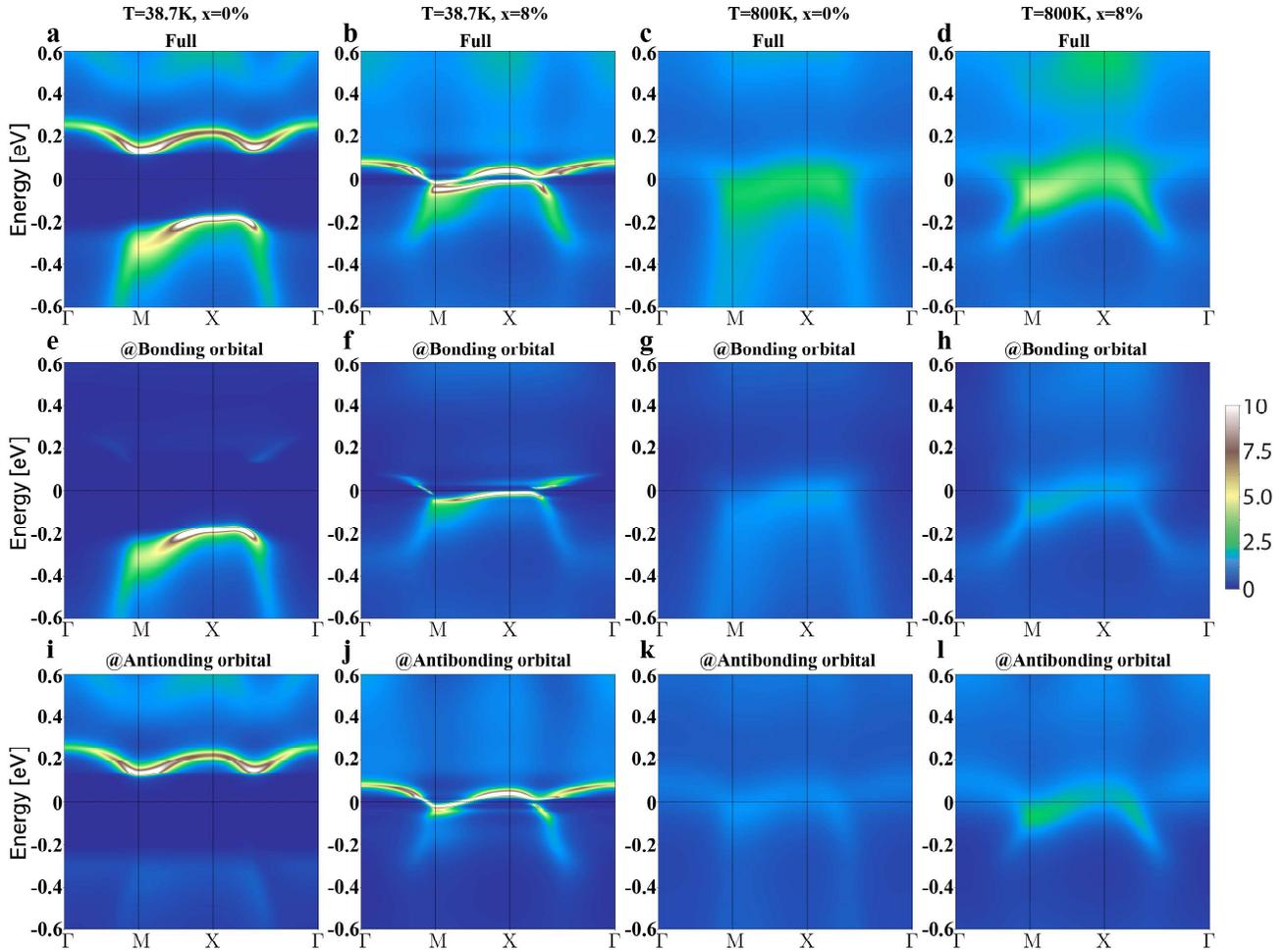

FIG. ST5. The orbital-resolved $A(\mathbf{k},\omega)$. The first and second column show the results of the undoped ($x=0\%$) and doped system ($x=8\%$), respectively, at $T=38.7$ K. The third and fourth column show the results of the undoped ($x=0\%$) and doped system ($x=8\%$), respectively, at $T=800$ K. The first row shows the full spectra, and the second (third) row shows the contribution to the full spectra from the bonding (anti-bonding) orbital.

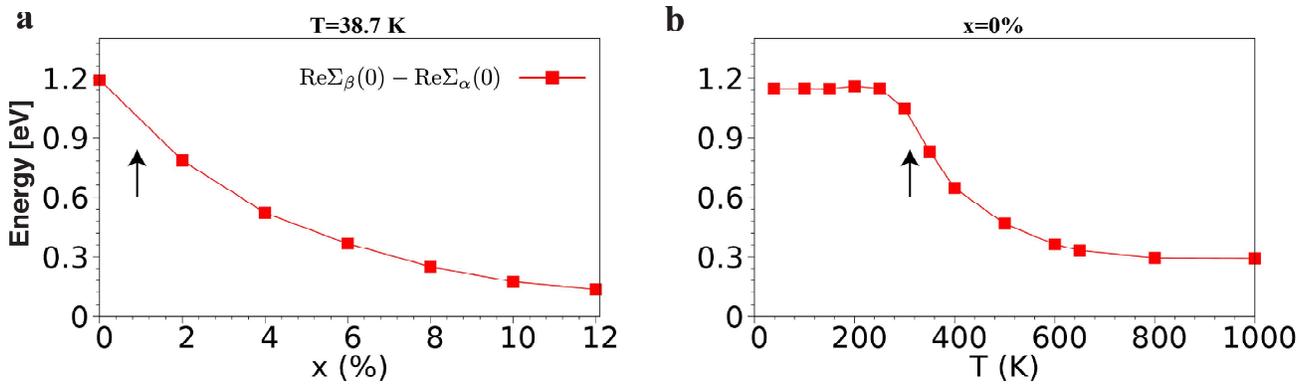

FIG. ST6. **a** (**b**) Doping (Temperature) dependence of $\mathrm{Re}\Sigma_\beta(0) - \mathrm{Re}\Sigma_\alpha(0)$. The arrows indicate a significant decrease in the magnitude of $\Sigma_\beta^R(0) - \mathrm{Re}\Sigma_\alpha(0)$ as one increases $x$ or $T$.

80% and 2% doping, as shown in Fig. ST11**a**. Consistent with this, the gap at the X-point does not disappear rapidly upon doping but instead decreases slowly between $x = 2\%$ and $x = 12\%$ doping, similar to the doping evolution of the antiferromagnetic correlations (Fig. ST11**a**).

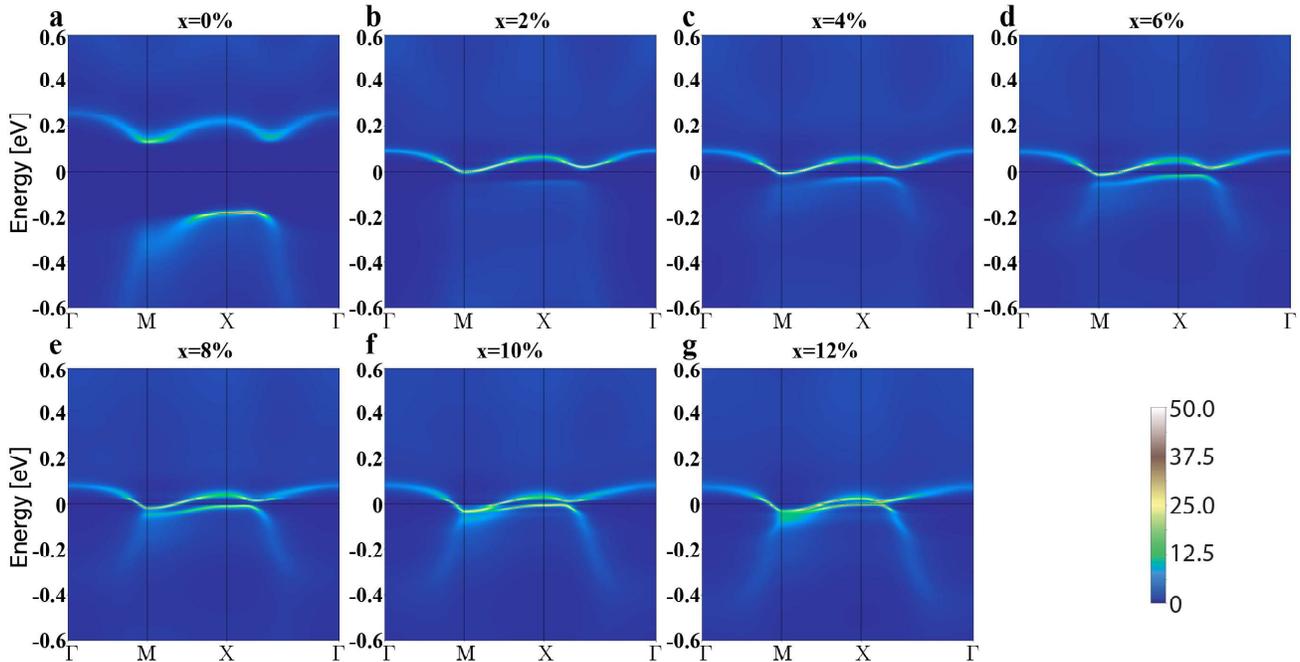

FIG. ST7. Momentum-resolved spetra along the $\Gamma$-X-$\Gamma$ **k**-path for the indicated doping levels ($x =$0%, 2% $\cdots$, 12%) at $T = 38.7$ K.

### E. High energy features of the spectral function

Figure ST8 shows $A(\mathbf{k}, \omega)$ in a larger energy window. The high-energy spectral features in the energy range $0.5 \lesssim E \lesssim 1$ eV are strongly broadened, consistent with the broad hump in the **k**-integrated spectra in Fig. ST4**a**. If we interpret this as the upper Hubbard band, the lower Hubbard band would be expected in the energy range $-1 \lesssim E \lesssim -0.5$ eV. In the undoped system, there is no clear indication for the presence of such a band, while a faint lower Hubbard band like feature becomes visible in the spectra of the doped systems.

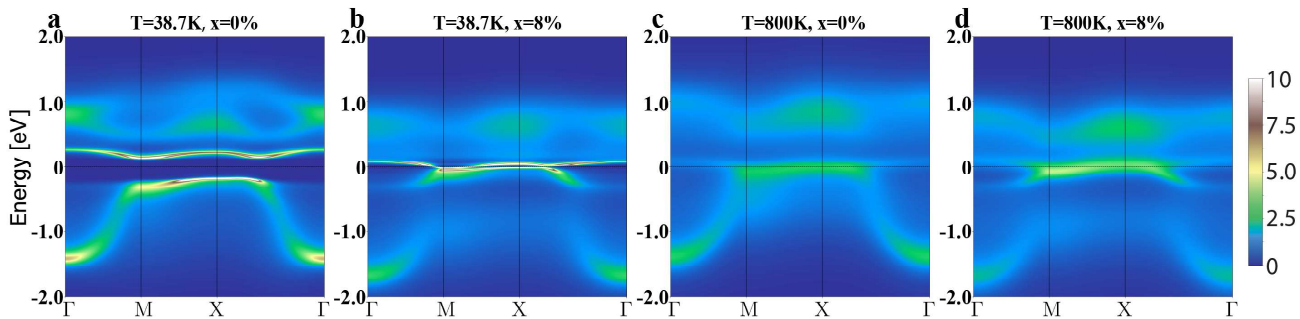

FIG. ST8. Momentum-resolved spetra along the $\Gamma$-X-$\Gamma$ **k**-path in the energy window $-2.0$ eV to $2.0$ eV.

### F. Constant energy cuts

Figure ST9 shows the evolution of the constant energy cuts with doping. At low-temperature, $T = 38.7$ K, the undoped system is gapped. As shown by panel **a**, there is no spectral weight for energies inside the gap but a large spectral weight is found at the X-point below at binding energy 0.2 eV, i.e., below the gap edge. As the doping concentration $x$ increases, the system becomes metallic and Fermi pockets appear and expand near the M point, which is consistent with the experimental results [1]. At high temperature and $x = 8\%$ doping (see Fig. ST5**d,h,i**), the bonding and anti-bonding bands merge and contribute similar spectral weight around the Fermi-energy (the anti-bonding orbital has slightly higher spectral weight). Above the Fermi energy, there is more spectral weight near the X point than near the M point and the bands at the M point are mainly below the Fermi energy.

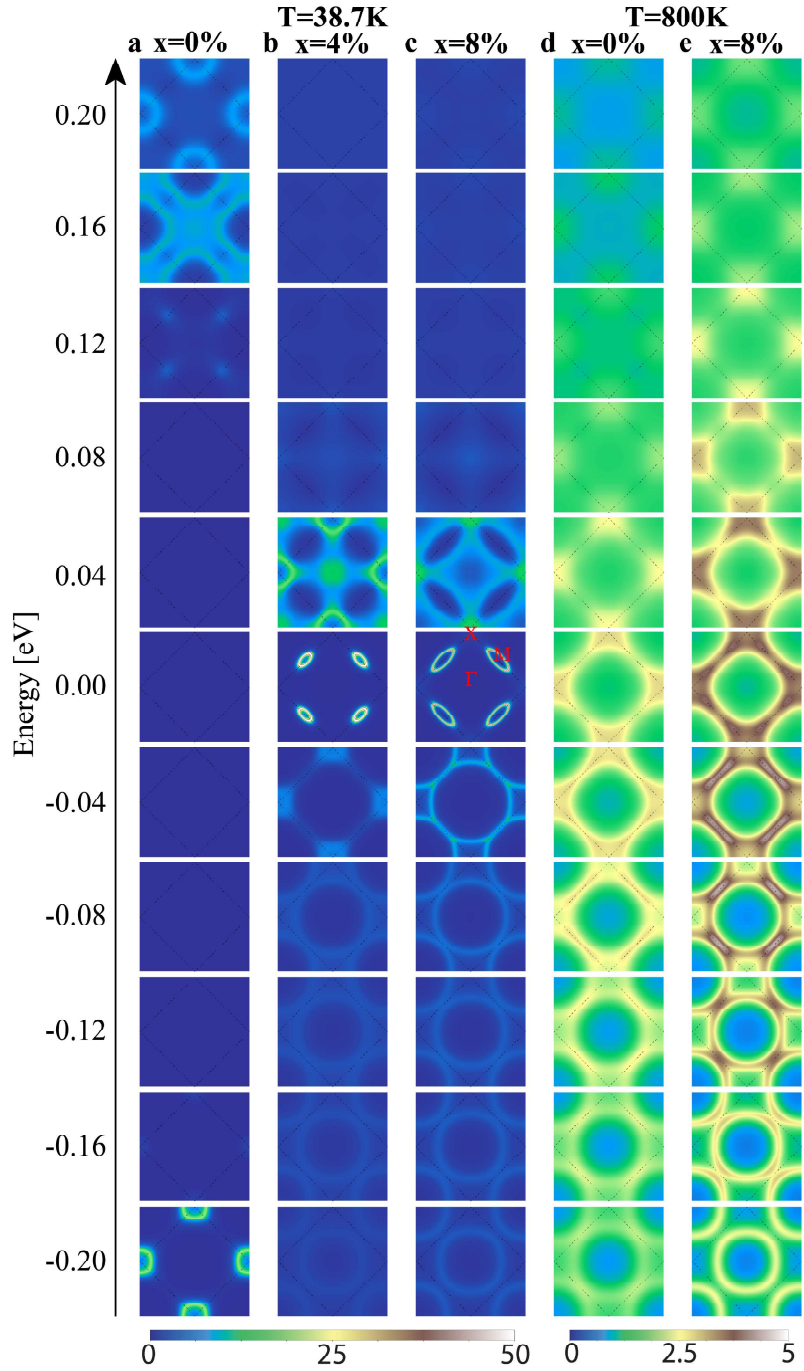

FIG. ST9. Constant energy cuts for the energies $-0.2, -0.16, \cdots, 0.2$ eV. The black dashed box in each panel illustrates the reduced Brillouin zone of the $\sqrt{2}\times\sqrt{2}$ unit-cell. **a-c** Results at temperature 38.7 K for the doping concentrations $x = 0\%$, 4%, and 8%, respectively. **d-e** Results at temperature 800 K for the doping concentrations $x = 0\%$ and 8%, respectively.

## G. Spin-Spin correlation functions on the cluster

The gap size of the undoped system at the M-point is roughly $E_g = 0.322$ eV at $T = 38.7$ K. If we regard this gap as a Mott gap, the melting (or filling in) of such a big gap requires a temperature $\sim O(11600 \text{ K/eV} * 0.322 \text{ eV}) = O(4000)$ K. However, as shown by Fig. ST4**d**, the gap closes at $T = 400$ K, which is 10 times smaller. As already argued above, the gap in $Sr_2IrO_4$ is unlikely of Mott origin. Instead, the gap is a correlated band insulator gap resulting from a correlation enhanced splitting of the bonding/antibonding bands. The gap closing can be related to the decreasing real part of the self-energy, as discussed above, and also to decreasing anti-ferromagnetic correlations as one increases $T$.

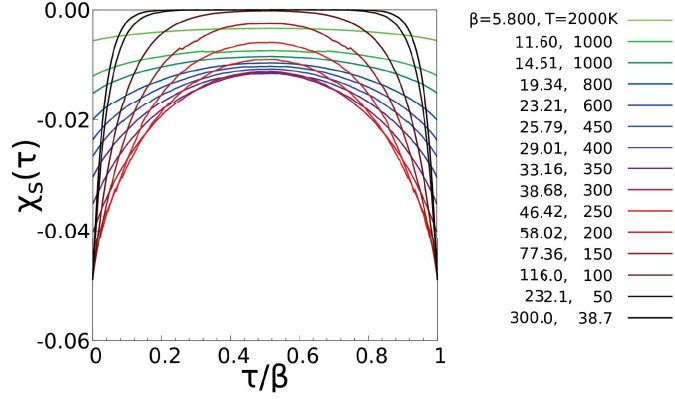

FIG. ST10. The spin-spin correlation function $\chi_s^{AB}(\tau)$ for indicated $T$ as a function of $\tau/\beta$.

The spin-operator for site $i$ reads $S_z^i = \frac{1}{2}(n_\uparrow^i - n_\downarrow^i)$, where the number operator at sites $A$ and $B$ in the same unit cell can be expressed in the bonding and anti-bonding basis as

$$n_\uparrow^A = \frac{1}{2}(n_{\alpha\uparrow} + n_{\beta\uparrow} + c^\dagger_{\alpha\uparrow} c_{\beta\uparrow} + c^\dagger_{\beta\uparrow} c_{\alpha\uparrow}), \tag{3}$$

$$n_\downarrow^A = \frac{1}{2}(n_{\alpha\downarrow} + n_{\beta\downarrow} + c^\dagger_{\alpha\downarrow} c_{\beta\downarrow} + c^\dagger_{\beta\downarrow} c_{\alpha\downarrow}), \tag{4}$$

$$n_\uparrow^B = \frac{1}{2}(n_{\alpha\uparrow} + n_{\beta\uparrow} - c^\dagger_{\alpha\uparrow} c_{\beta\uparrow} - c^\dagger_{\beta\uparrow} c_{\alpha\uparrow}), \tag{5}$$

$$n_\downarrow^B = \frac{1}{2}(n_{\alpha\downarrow} + n_{\beta\downarrow} - c^\dagger_{\alpha\downarrow} c_{\beta\downarrow} - c^\dagger_{\beta\downarrow} c_{\alpha\downarrow}). \tag{6}$$

The imaginary-time AFM spin-spin correlation function on the cluster is defined as $\chi_s^{AB}(\tau) = \langle S_z^A(\tau) S_z^B(0) \rangle$ and can be measured in CT-HYB. If $\chi_s^{AB}(\tau) < 0$, we have anti-ferromagnetic spin correlations within the cluster.

Figure ST10 shows the evolution of $\chi_s^{AB}(\tau)$ as one increases $T$ in the half-filled system. The value of $\chi_s^{AB}(\tau)$ is negative on the whole imaginary time ($\tau$) axis, which shows that the system has short-ranged AFM correlations. The rapid decay of $\chi_s^{AB}(\tau)$ at low $T$ (black curve in Fig. ST10) indicates that a strong singlet is formed on the two-site cluster. A singlet state yields large AFM correlations at $\tau = 0$, but the correlations as a function of time decay fast because the spin flips up and down under the time evolution. The singlet formation can thus be detected by tracking the evolution of $\chi_s^{AB}(\beta/2)$.

In Fig. ST11**a,b** we show the evolution of the instantaneous AFM correlations $\chi_s^{AB}(\tau = 0)$ with doping and temperatue, and in panels **c,d** the AFM spin susceptibility $\chi_s \equiv \int_0^\beta \chi_s^{AB}(\tau) d\tau$ together with the long-time value $\beta \chi_s^{AB}(\beta/2)$ of the correlation function. One can see that the dependence of both $\chi_s^{AB}(\tau = 0)$ and $\chi_s$ on temperature and doping is similar to the evolution of the correlation enhanced level splitting in Fig. ST6 and to the evolution of the gap in the spectrum (Fig. ST5), even though the AFM short-range correlations do not vanish at the gap closing point. The relative drop in the local susceptibility is larger than that in the instantaneous correlations. It is interesting to note that as a function of temperature, strong singlet formation sets in below $T \approx 300$ K, which is the same temperature where the correlation-induced level splitting and the gap size saturate. The weak temperature dependence of the gap in $A(\mathbf{k}, \omega)$ up to 300 K can thus be linked to the presence of a strong singlet on the cluster.

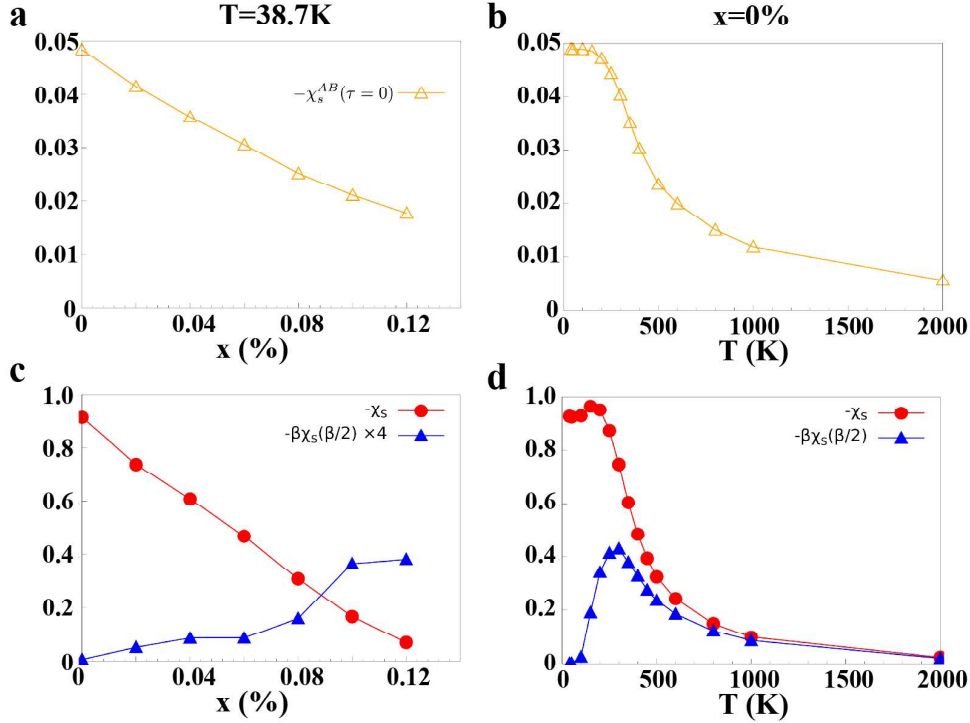

FIG. ST11. **a,c** (**b,d**) Doping ($T$) dependence of the anti-ferromagnetic correlations at fixed $T = 38.7$K (in the undoped system).

## ST2. GW+EDMFT CALCULATIONS

In order to address the question of dynamical screening we employ GW + extended dynamical mean-field theory (GW+EDMFT) [17–20] simulations of the extended Hubbard model, using the causal self-consistency scheme [15, 16]. Compared with the model in Sec. ST1 A, we include here the effect of a non-local Coulomb interaction. The calculations are performed at different temperatures with a bare onsite repulsion $U = 1.5$ eV and fairly large nearest neighbor repulsions $V$ up to 0.5 eV.

### A. Effective local interaction

GW+EDMFT employs a self-consistently computed frequency-dependent onsite interaction $\mathcal{U}(\omega)$ in the impurity model and thus allows to capture the screening from local and nonlocal charge fluctuations. The frequency dependence on the Matsubara axis is plotted for different temperatures, and for the largest nonlocal interaction ($V = 0.05$ eV) in Fig. ST12. These data show that at low temperature, the static screened interaction $\mathcal{U}(\omega = 0)$ is about 10% lower than the bare $U$, while at high temperature, the screening reduces the static interaction by about 30%. For $V = 0.3$, the screening is less than 10% even at high temperature. A screening effect of similar magnitude is found for 8%

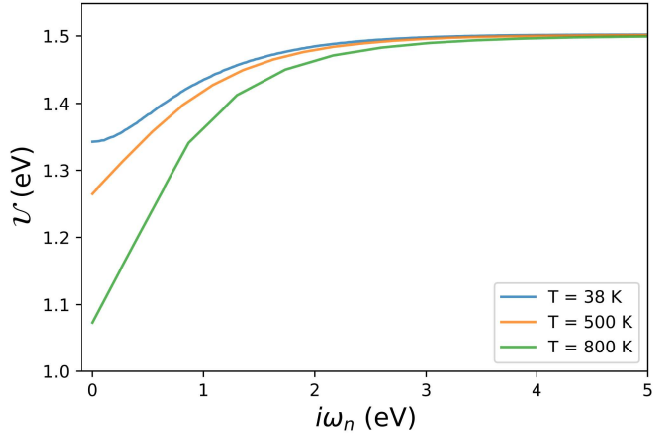

FIG. ST12. Matsubara-frequency dependence of the effective local interaction, $\mathcal{U}(i\omega_n)$, for the half-filled system with bare interaction $U = 1.5$ eV and nearest-neighbor interaction $V = 0.5$ eV for indicated temperatures.

doping, see Tab. ST1.

We consider a single-site impurity problem in the EDMFT calculation, but allow for AFM long-range order by implementing the self-consistency loop for a two-site unit cell. At half-filling, the system becomes an anti-ferromagnetic (AFM) insulator below a Néel temperature of about 550 K, as shown in Fig. ST13. Panel **a** plots spin averaged spectra for $T = 38, 500, 600$ and $800$ K, while panel **b** shows the staggered magnetization as a function of temperature. Figure ST14 shows that the gap in the low-temperature solution disappears if the calculation is restricted to the PM phase. This demonstrates that AFM order is (within GW+(single-site)EDMFT) necessary to open a gap.

TABLE ST1. Temperature dependence of the screened effective local interaction $\mathcal{U}(\omega = 0)$ for the half-filled and the 8% doped system, with the bare interaction fixed at $U$=1.5.

|       | $x = 0\%$ | $x = 0\%$ | $x = 8\%$ |
| ----- | --------- | --------- | --------- |
| $T$ (K) | $V = 0.3$ | $V = 0.5$ | $V = 0.3$ |
| 38.7  | 1.48      | 1.34      | 1.34      |
| 500   | 1.46      | 1.27      | 1.38      |
| 600   | 1.44      | 1.05      | 1.39      |
| 700   | 1.42      | 1.02      | 1.40      |
| 800   | 1.40      | 1.07      | 1.40      |

### B. Momentum-resolved spectra

Finally, we plot in Fig. ST13 the momentum-resolved spectral functions along the path Γ-M-X-Γ for the low-temperature undoped system (panel **a**) and for the high-temperature 8% doped system (panel **b**). These spectra can be compared with the PM OC-DMFT results in Fig. ST5. We find that both OC-DMFT (with only short-ranged AFM correlations) and GW+EDMFT (with long-ranged AFM order) provide a consistent description of the temperature and doping dependent electronic structure of $Sr_2IrO_4$. In particular both capture the opening of a gap and a correlated band insulator state at low $T$ in the half-filled system and the emergence of a broadened, but only weakly renormalized metallic band in the doped high-$T$ system.

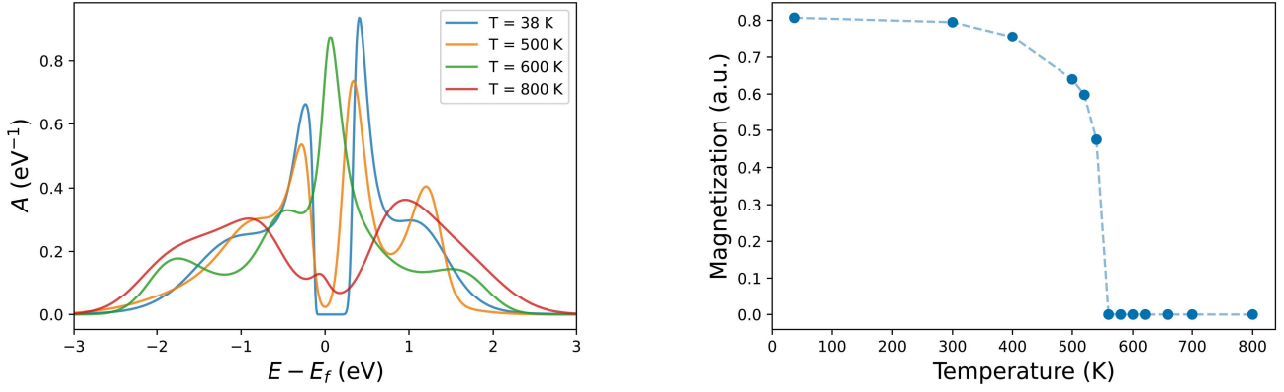

FIG. ST13. Left panel: Spin averaged local spectral function ($A(\omega)$) of half-filled $Sr_2IrO_4$ at different temperatures. Right panel: Staggered magnetization of half-filled $Sr_2IrO_4$ as a function of temperature. The AFM order disappears around 550 K, where the insulator-metal phase transition occurs.

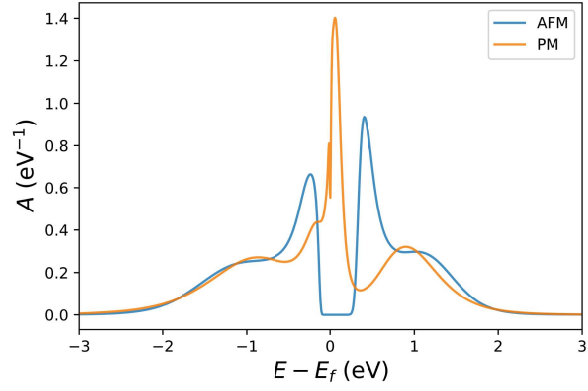

FIG. ST14. Comparison of the local spectral function $A(\omega)$ of the AFM and PM system at $T = 38.7$ K.

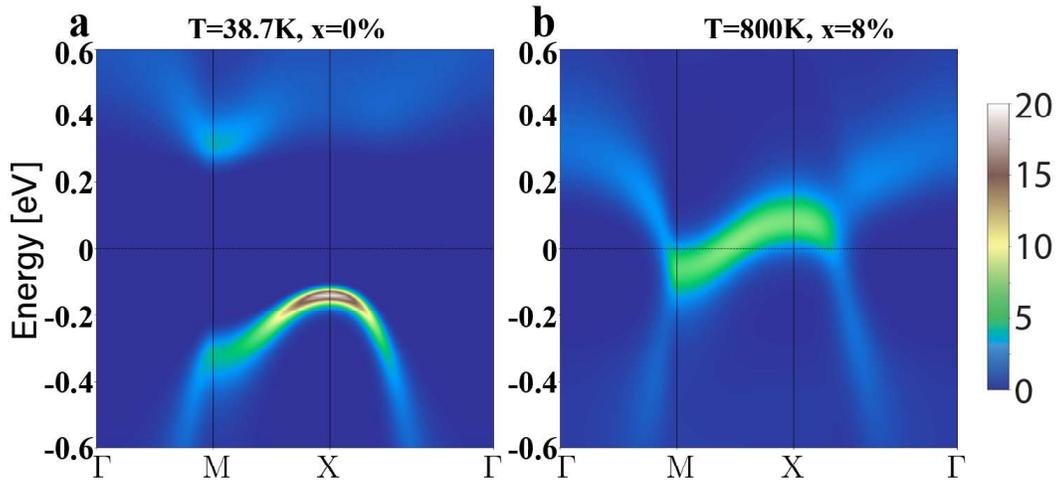

FIG. ST15. Momentum-resolved spetra obtained by the GW+EDMFT method. **a** Results at $T = 38.7$ K for the undoped ($x = 0\%$) system. **b** Results at $T = 800$ K for the doped ($x = 8\%$) system.

Supplementary Information (experimental part) for

**Light-induced insulator-metal transition in $Sr_2IrO_4$ reveals the nature of the insulating ground state**

# SE1. Energy distribution curve at the X point in a static spectrum

To check the consistency of our spectra with those from the literature, we extracted an energy distribution curve (EDC) at the X point in a static spectrum and performed a fitting on it as shown in Fig. SE1c. The analytical form of the fitting function is as follows.

$$I(E) = [(SG_1 + SG_2 + bkg) \otimes G](E) \quad (1)$$

$$SG_i(E) = A_i \times \exp\left(-\left(\frac{E - E_i}{\frac{w_i}{2\sqrt{\ln(2)}}}\right)^2\right) \times \left(1 - \mathrm{erf}\left(\alpha_i \times \frac{E - E_i}{\frac{w_i}{2\sqrt{\ln(2)}}}\right)\right) \quad (2)$$

where SG(E) is a skewed Gaussian function for a peak function, bkg(E) is a Shirley function for a background, $\otimes$ is the convolution, G(E) is a Gaussian function, and erf() is the error function. Note that α in Eq. (2) controls the skewness of the Gaussian function. Although the detailed functional form may be different, skewed Gaussian functions, or asymmetric Gaussian functions, were also used for fitting EDCs from pristine and Ru-doped $Sr_2IrO_4$ in ref. [1]. The energy resolution, 0.036 eV, extracted from $Bi_2Se_3$ spectra (see SE7) was used in the convolution with the Gaussian function. The Fermi-Dirac distribution was not used since at T ~ 37.4 K, $Sr_2IrO_4$ is known to be in an insulating state with a gap much bigger than $k_B T$. In Fig. SE1c, the fitting range is indicated by the two red-dotted vertical lines, and $SG_1(E)$ and $SG_2(E)$ are noted as peak1 and peak2, respectively. The peak locations of $SG_1(E)$ and $SG_2(E)$ are -0.11 eV and -0.24 eV, respectively, which are consistent with those at the X point from the static spectrum of undoped $Sr_2IrO_4$ in ref. [2] (~ -0.10 eV and ~ -0.26 eV, respectively).

Note that the calculated spectrum shows only one peak at the X point (Fig. 1f). Our simulation does not include $J_{eff}$ = 3/2 states (see ST1.A.), but, with all six bands of $t_{2g}$ character, at the X point, the next band is more than 0.5 eV below the valence band according to the tight-binding calculation in both calculations with U = 0 eV and 2 eV in ref. [2]. Currently, the reason why two peaks appear with a close distance to each other near $E_F$ at the X point in the experimental results (Fig. SE1c) and whether one of these features is associated with $J_{eff}$ = 3/2 states are unclear. In ref. [2], at the X point of undoped $Sr_2IrO_4$, they observed two peaks which are extracted from two- or one-dimensional momentum curvature plots[3] and they assigned the valence band to the peak at ~ -0.10 eV (closer to $E_F$), which coincides with the overall landscape of peaks at the M and Γ points. However, several other studies summarized in ref. [4] show a single peak at ~ -0.25 eV at the X point. In our simulation in Fig. 1f, the location of the valence band ($J_{eff}$ = 1/2 band) maximum is set at $E - E_F \approx$ -0.18 eV which is the middle point of the locations of the two peaks in Fig. SE1c (-0.11 eV and -0.24 eV). In principle, $E_F$ can be tuned in the simulation, so the essential physics does not depend on whether the valence band maximum in the simulation is at -0.11 eV or -0.24 eV or a value in between.

The expected energy locations of the conduction band at the branch and X point are ~ 0.15 eV and ~ 0.22 eV, respectively, from Fig. 3d, Fig. SE5d, and Fig. ST3a. If we consider the distance between the valence band maximum in the simulation (~ -0.18 eV, Fig. 1f) and the peak locations (-0.11 eV and -0.24 eV) in Fig. SE1c, the expected energy locations of the conduction band in our data are 0.09 eV or 0.22 eV at the branch and 0.16 eV or 0.29 eV at the X point. (In the main text, we mentioned that the conduction bands are expected to be located in the range of 0.1 – 0.2 eV at the branch and 0.15 – 0.3 eV at the X point.) Those values are close to the estimated energy locations of the conduction band at the branch and X points of ~ 0.18 eV and ~ 0.26 eV, respectively, from the tight-binding calculation in ref. [2]. Here, the energy location of the conduction band at the branch in ref. [2] is estimated as follows: The conduction band location at the M point is ~ 0.15 eV. The position of the branch in momentum space in our spectra is shifted by 1/4 of the X – M from the M point (Fig. 3a). If for simplicity we assume that the energy location of the conduction band changes linearly from the M point to the X point, then it will be ~ 0.18 eV at the branch in the tight-binding calculation in ref. [2].

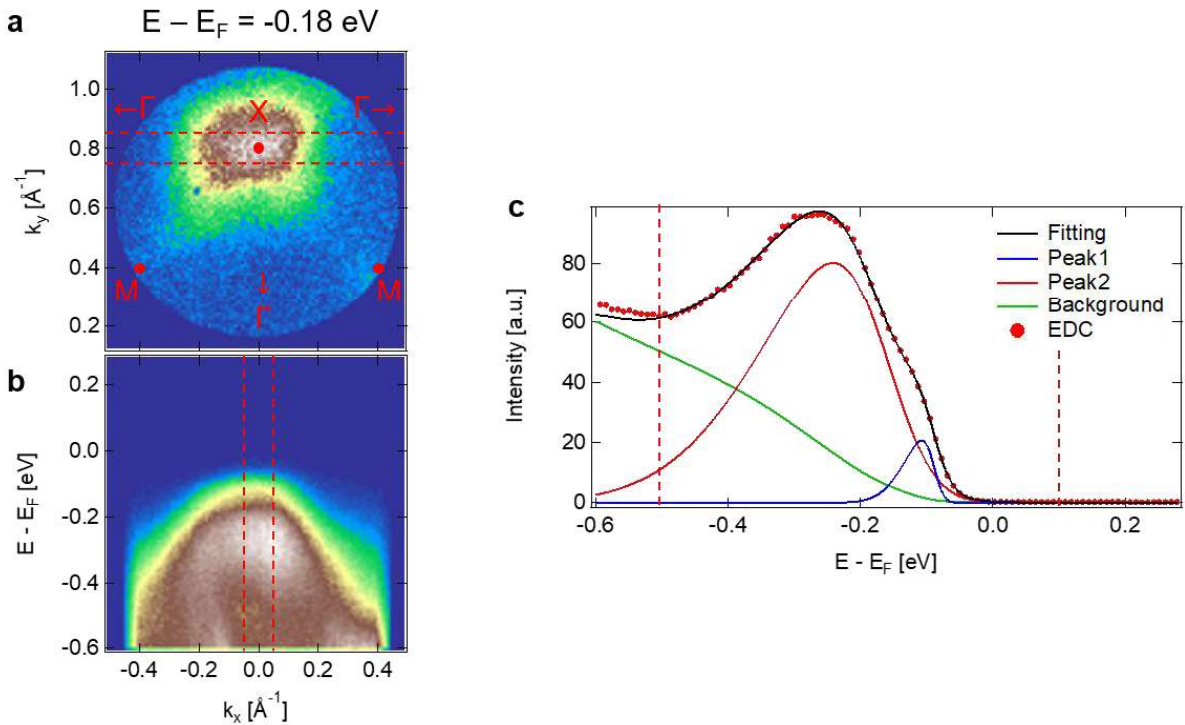

**Fig. SE1. Static ARPES spectrum of $Sr_2IrO_4$ and the fittings on the EDC at the X point. a**, A constant energy cut at $E - E_F$ = -0.18 eV from a static ARPES spectrum of $Sr_2IrO_4$. The high symmetry points (the X and M points) and directions towards the Γ points are marked. **b**, E-k cut along the $\Gamma - X - \Gamma$ direction. The slicing location and the integration range are indicated by the pair of red dotted lines in **a**. Note that **a** and **b** are the same as Figs. 1c and e. **c**, EDC at the X point. The slicing location and the integration range are indicated in **b** by the pair of red dotted lines. Fitting with two skewed Gaussian peaks[1] is performed on the EDC in the range marked by the two red dotted lines. Peak1 is the origin of the hump at $E - E_F \approx$ -0.1 eV. The energy locations of the maxima of Peak1 (-0.11 eV) and Peak2 (-0.24 eV) match those at the X point from the static spectrum of undoped $Sr_2IrO_4$ in ref. [2].

**SE2. The location and range of the momentum integration for the EDCs at the branch and the X point**

In Fig. SE2a and b, the red boxes show the locations and ranges of the momentum integration for the EDCs at the branch and the X point, respectively. ($\Delta k_x$, $\Delta k_y$) for the integration range at the branch and the X points are (0.3 Å$^{-1}$, 0.1 Å$^{-1}$) and (0.1 Å$^{-1}$, 0.1 Å$^{-1}$), respectively. The reason for the wide integration range is not to miss relevant spectral weight, especially any possible signatures of the conduction band, at different energies since the center of the spectrum slightly moves as energy increases, which is shown in the plots for the peak position from the Gaussian fitting of MDCs as a function of energy (Figs. SE6f and 7f).

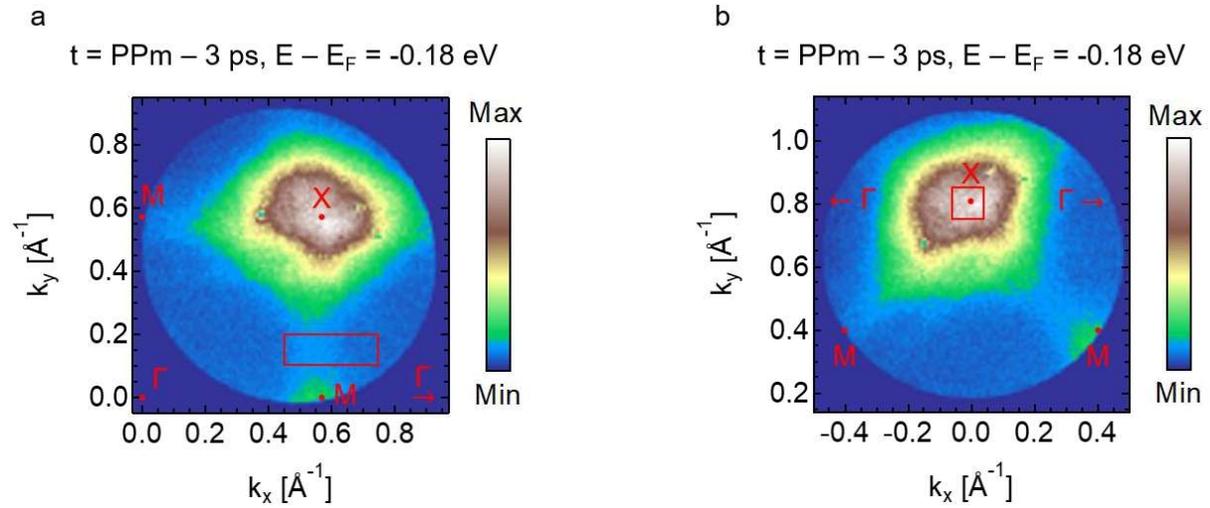

**Fig. SE2. The location and range of the momentum integration for the EDCs at the branch and the X point. a, b,** Constant energy cuts at $E - E_F$ = -0.18 eV at PPm – 3 ps. The location and range of the momentum integration for the EDCs at the branch and the X point are indicated by the red boxes in **a** and **b**, respectively. The high symmetry points (the Γ, X, and M points) and the direction towards the Γ point are marked. Note that the color scale in **a** and **b** is linear and ranges from the minimum value to the maximum value of the intensity in each panel.

## SE3. Estimation of electronic temperature

When an insulating material is perturbed by a pump pulse with photon energy above the band gap, this naturally leads to an increase in the electronic temperature and photo-doping. In this section, we describe how the electronic temperature was estimated from our data. After the absorption of the pump beam, the initial population of the excited state will be non-thermal. Since thermalization by electron-electron scattering can take few hundred femto-seconds, at PPm which is ~140 fs after the temporal overlap, the electronic system has not yet thermalized. This means the temperature of the system cannot be defined and an EDC at the X point cannot be fitted by the Fermi-Dirac distribution[5]. On the other hand, at PPm + 1 ps, the Fermi-Dirac distribution can be used to fit an EDC at the X point as shown in Fig. SE3a and the extracted electronic temperature is 434 $\pm$ 9 K, which means the electronic system is (almost) thermalized by the electron-electron scattering. The fitting range was taken up to 0.1 eV where the slope of the EDC in the semi-log plot changes. Above 0.1 eV, the slope stays nearly constant up to high energies. It is expected to be a background instead of a signal because a similar background signal also appears as a constant slope to higher energies in PPm − 3 ps as shown in Fig. 2b and c. At PPm, the system cannot be described by a single electronic temperature. Instead, we can try to extract an energy dependent effective temperature[6]. Three fittings with the Fermi-Dirac distribution are applied in three different energy ranges, lower (-0.18 to -0.05 eV), middle (-0.05 to +0.05 eV), and upper (+0.1 to +0.3 eV) bands. Within each range, the density of states can be reasonably approximated by constants from our calculation shown in Fig. SE3b. As shown in Fig. SE3c, since the slope of the EDC does not become a constant up to 0.3 eV, we expect that the data is a meaningful signal instead of a background (at least) up to 0.3 eV. In Fig. SE3c, 456 ($\pm$ 16), 818 ($\pm$ 138), and 1927 ($\pm$ 218) K are extracted by fitting the Fermi-Dirac distribution functions on the lower, middle, and upper bands, respectively. The analytical form of the fitting function is as follows.

$$f(E) = \frac{A}{1 + \exp\left(\frac{E - E_F}{k_B T}\right)} \quad (3)$$

where A is a constant density of states, $E_F$ is the Fermi level, $k_B$ is the Boltzmann constant and T is the temperature. It makes sense to have higher electronic temperatures at higher energies with smaller populations (e.g. the middle and upper bands) and they relax to lower temperature reservoirs with larger populations at lower energies (e.g. the lower band). This result indicates that, at PPm, the electronic temperature ranges from ~400 to ~2000 K. Note that, in the middle and upper bands, the fitting uses the high energy tail of the Fermi-Dirac distribution and there is no information on the density of states (A) and the inflection point at E = $E_F$ in the data. This allows many different sets of the converged fit parameters, (A, $E_F$, T). Here the convergence of the fitting is based on a certain criterion of Chi-square. In practice, we selected a fit parameter set (A, $E_F$, T) which yields a converging fit with a reasonably small absolute $E_F$ value. The T value does not change much in the different fit parameter sets (A, $E_F$, T) with $E_F$ down to ~ -1.0 eV. The selected values of the fit parameter set, (A, $E_F$, T), are listed in Table S1. For the simulation of the E-k cut in Fig. 3e, we used 800 K, which is close to the value from the middle band, as a representative temperature of the system for the calculation. We did not pick up the temperature from the upper band since the intensity which is proportional to the electron population is too small.

Table SE1. The fitting parameters of the Fermi-Dirac distribution on the EDC at PPm

|  | Lower band | Middle band | Upper band |
|---|---|---|---|
| A [a.u.] | 489.28 ± 12.6 | 2816.3 ± 2.99×10$^4$ | 272.82 ± 1.1×10$^3$ |
| $E_F$ [eV] | -0.11982 ± 0.00246 | -0.30964 ± 0.81 | -0.41975 ± 0.754 |
| T [K] | 455.72 ± 15.8 | 817.52 ± 138 | 1927.1 ± 218 |

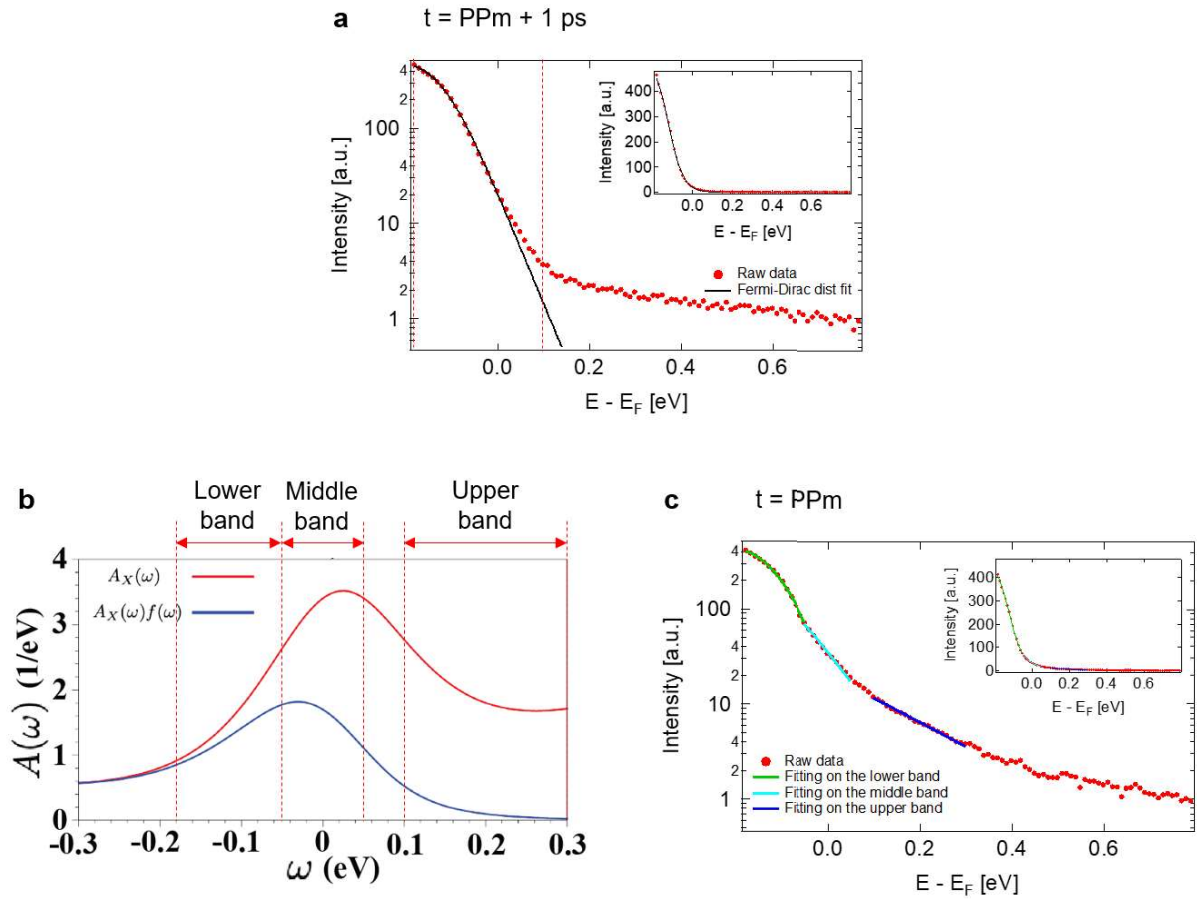

**Fig. SE3. Estimation of the electronic temperature. a**, A fitting with the Fermi-Dirac distribution on the EDC at the X point at PPm + 1 ps. The Fermi-Dirac distribution well fits the EDC, which means the system is essentially thermalized and the extracted temperature is 434 ± 9 K. **b**, The calculated density of states at the X point (without momentum integration), $A_X(\omega)$, and the Fermi-Dirac distribution, $f(\omega)$, at the temperature of 800 K and the electron doping of 8% electrons/Ir-site. The energy ranges for the lower, middle, and upper bands are shown and we consider the density of states to be approximately constant within each range. **c**, Fittings with the Fermi-Dirac distribution on the lower, middle, and upper bands of the EDC at the X point at PPm. Note that the location and range of the momentum integration for the EDCs at the X point are indicated by the red box in Fig. SE2b.

## SE4. Estimation of the photo-doping level

In this section, we describe how the photo-doping level was estimated in our data. The simplified expression for calculating the photo-doping level is as follows[7].

$$d = \frac{N}{n \times A \times \xi} \qquad (5)$$

where d is the photo-doping level [#photons/Ir-site], N is the number of photons per pump pulse, n is the number of Ir sites per unit volume, A is the area illuminated by the pump beam on the sample surface, and $\xi$ is the optical penetration depth of photons. We assumed that one incident photon generates one excited electron (conversion efficiency = 100%). As the pump beam penetrates the material, the photo-doping level decreases as a function of distance along the penetration direction. Also, since only photo-emitted electrons very close to the surface can escape the material, the depth or electron mean free path ($z_{pr}$) to be considered is very small. Let us consider the situation shown in Fig. SE4a where the pump beam is refracted with an angle $\theta_t$ and it penetrates the material along the direction, l. If a thin disk at position l is considered, the number of photons absorbed in the disk is the difference between the number of incident and exiting photons. Since N is proportional to the power of the pump beam, P, which is a function of l, the number of photons per pump purse in the disk, $\Delta N(l)$, can be formulated as follows.

$$\Delta N(l) = N(l) - N(l + \Delta l) = \frac{N(0)}{P(0)} \times \frac{[P(l) - P(l + \Delta l)]}{\Delta l} \times \Delta l \qquad (6)$$

where $\Delta l$ is the thickness of the disk along the penetration direction. The photo-doping level at z (in the vertical direction) can be expressed as follows.

$$d(z) = \frac{\Delta N(l(z))}{n \times A \times \Delta z} = \frac{1}{n \times A} \times \frac{N(0)}{P(0)} \times \frac{[P(l) - P(l + \Delta l)]}{\Delta l} \times \frac{1}{\cos(\theta_t)} \qquad (7)$$

where $l(z) = z/\cos(\theta_t)$. By Snell's law, $\cos(\theta_t)$ can be expressed as follows.

$$\cos(\theta_t) = \sqrt{1 - \left(\frac{n_i}{n_t}\right)^2 \times \sin^2(\theta_i)} \qquad (8)$$

where $n_i$ and $n_t$ are refractive indexes of the vacuum and the sample, respectively and $\theta_i$ is the incidence angle of the pump beam on the sample surface. If we take a limit of $\Delta l \to 0$, then d(z) becomes as follows.

$$d(z) = \frac{N(0)}{n \times A} \times \left[\frac{1}{P(0)} \times \frac{1}{\cos(\theta_t)} \times \left(-\frac{dP(l)}{dl}\right)\right] \qquad (9)$$

From Beer-Lambert's law, P(l) can be expressed as follows.

$$P(l) = P(0) \times \exp(-\alpha l) \tag{10}$$

where α is an absorption coefficient. Using this, we can express d(z) as follows.

$$d(z) = \frac{N(0)}{n \times A} \times \left[\frac{\alpha}{\cos(\theta_t)} \times \exp(-\alpha l)\right] \tag{11}$$

To evaluate an average photo-doping level from z = 0 to $z_{pr}$, we can consider $d_{avg}$.

$$d_{avg} \times z_{pr} = \int_0^{z_{pr}} d(z) dz \tag{12}$$

Then, we get $d_{avg}$ as follows.

$$d_{avg} = \frac{N(0)}{n \times A \times \xi \times \cos(\theta_t)} \times \left[\frac{1}{\alpha l_{pr}}\left(1 - \exp(-\alpha l_{pr})\right)\right] \tag{13}$$

where $\xi = 1/\alpha$, $l_{pr} = z_{pr}/\cos(\theta_t)$. In the following sections, we describe how we calculated each factor.

SE4.1. Estimation of the number of photons per pump pulse at the sample surface, N(0)

The number of photons per pump pulse at the sample surface, N(0), can be estimated by the following equation.

$$N(0) = t_w \times (1 - R) \times \frac{P_{in}}{f \times E_{ph}} \tag{14}$$

where $t_w$ is the transmission rate of the window (uncoated ZnSe), R is the reflectivity at the surface of the sample, $P_{in}$ is the power of the incidence pump beam on the ZnSe window, f is the repetition rate of the pump pulse, $E_{ph}$ is the photon energy of the pump pulse. We used $t_w$ = 0.70, R = 0.24, $P_{in}$ = 30.3 mW, f = 300 kHz, $E_{ph}$ = 1.204 eV. $t_w$ value was taken from the transmission curve provided by Thorlabs, Inc. for 5 mm thick uncoated ZnSe window[8]. The thickness of our window is ~2.5 mm and we assumed $t_w$ is not changed significantly for thickness from 5 mm to ~2.5 mm since this is the case for the wavelength of above 2.5 um (to ~13 um) according to the transmission curve provided by Crystran Ltd.[9]. R value was calculated through the Fresnel equations at the interface of the vacuum and an absorbing media[10] and for the calculation, we used the values of permittivity and optical conductivity in ref. [11] (Also, see SE4.4.). The calculated value of N(0) is ~2.8 × $10^{11}$ [#photons / pump pulse].

SE4.2. Estimation of the number of Ir sites per unit volume, n

The number of Ir sites per unit volume can be estimated by the following equation,

$$n = \frac{\#\text{Ir sites}}{a \times b \times c} \quad (15)$$

where a, b, and c are the two in-plane constants and the out-of-plane constant of a unit cell, respectively. For $Sr_2IrO_4$, the unit cell has the lattice constants of a = b ≅ 5.49 Å and c ≅ 25.78 Å for the in-plane and out-of-plane components, respectively[12]. It is a $\sqrt{2} \times \sqrt{2} \times 2$ supercell of the in-plane square lattice with the lattice constant of $a_0 = a/\sqrt{2}$ ≅ 3.88 Å[13]. Also, there are eight iridium atoms in a unit cell. The calculated value of n is ~$1.0 \times 10^{28}$ [#Ir sites / $m^3$].

SE4.3. Estimation of the area illuminated by the pump beam on the sample surface, A

We tilted the sample in one direction to reach the X point. The pump beam is injected with an angle of ~46.1° from the sample normal without tilting. The area illuminated by the pump beam on the sample surface, A, can be estimated by the following equation.

$$A = \pi \times \frac{D_1}{2} \times \frac{D_2}{2} \times \frac{1}{\cos((90° - 46.1°) - \theta)} \quad (16)$$

where $D_1$ and $D_2$ are the major and minor diameters of the pump beam and θ is the tilt angle of the sample. The sample was tilted towards the pump beam. The schematic is shown in Fig. SE4b. We approximately placed the focal spot of the pump beam at the sample surface by adjusting the focusing lens position. The major and minor diameters of the beam spot at the focal plane are 80 μm and 73 μm, respectively, based on $1/e^2$ beam size. We tilted the sample by $\theta$ ≅ 13.2°. The calculated value of A is ~$5.3 \times 10^{-9}$ [$m^2$].

SE4.4. Estimation of the optical penetration depth of the photons, ξ

Let us use the following notations for the complex refractive index (Eq. (17)), complex permittivity (Eq. (18)), and complex conductivity (Eq. (19)).

$$\mathbf{N} = n + i\kappa \quad (17)$$

$$\boldsymbol{\epsilon} = \epsilon_1 + i\epsilon_2 \quad (18)$$

$$\boldsymbol{\sigma} = \sigma_1 + i\sigma_2 \quad (19)$$

The optical penetration depth of the photons, ξ, can be estimated by using the following relationships among the optical properties[7,14].

$$\xi = \frac{\lambda}{4\pi\kappa} \quad (20)$$

$$\kappa = \sqrt{\frac{\sqrt{\epsilon_1^2 + \epsilon_2^2} - \epsilon_1}{2}} \quad (21)$$

$$\epsilon_2 = \frac{\sigma_1}{\omega} \quad (22)$$

where $\lambda$ is the wavelength of the pump beam in vacuum and $\omega$ is the angular frequency of the pump beam. From ref. [11] for single crystal $Sr_2IrO_4$ samples, we digitized the plots and interpolated the values, and then we got $\epsilon_1$ = 2.3 and $\sigma_1$ = 771 [$\Omega^{-1} \cdot cm^{-1}$] for the photon energy of 1.204 eV. From the above relationships, we calculated that $\epsilon_2$ = 4.8, $\kappa$ = 1.2, and $\xi$ = 67 nm. This value of the penetration depth is consistent with the values from other literatures[15,16].

SE4.5. Calculation of the photo-doping level, $d_{avg}$

$n_t$ is calculated from the following relation,

$$n = \sqrt{\frac{\sqrt{\epsilon_1^2 + \epsilon_2^2} + \epsilon_1}{2}} \quad (23)$$

With $\epsilon_1$ = 2.3 and $\epsilon_2$ = 4.8, we get $n_t$ = 1.9. The calculated value of $\cos(\theta_t)$ is 0.96 from Eq. (8) with $n_i$ = 1, $\theta_i$ = 32.9°. If an electron at the Fermi level is considered, its kinetic energy is 17.4 eV, corresponding to the difference between the photon energy of the probe beam and the work function of the analyzer (21.6 − 4.2 eV)[17]. Then its mean free path, $z_{pr}$, is ~0.7 nm[18]. With the values of $\alpha$ (= 1/$\xi$) and $l_{pr}$ (= $z_{pr}$/cos($\theta_t$)), $\left[\frac{1}{\alpha l_{pr}}(1 - \exp(-\alpha l_{pr}))\right]$ = 0.995. From Eq. (13), the photo-doping level, $d_{avg}$ = ~0.079 = 7.9% [electrons/Ir-site].

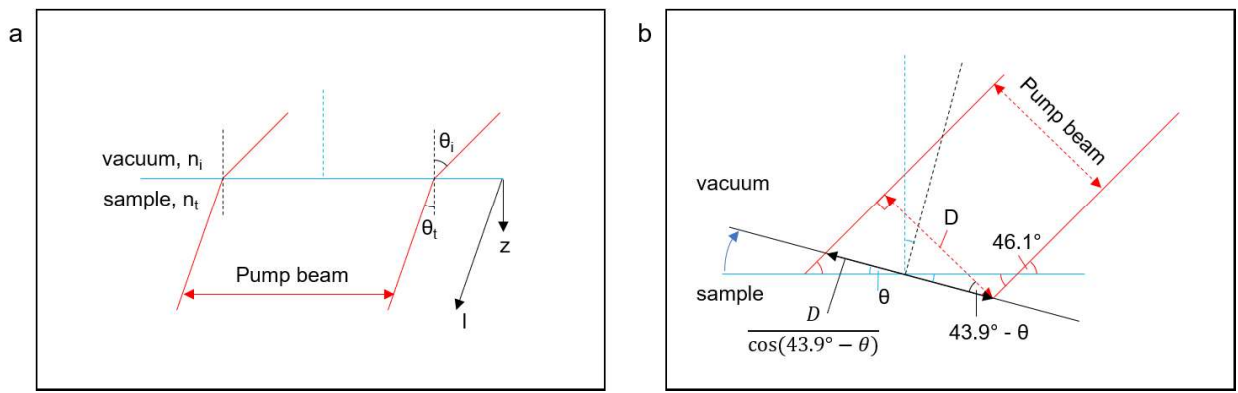

**Fig. SE4. Schematics of the pump beam at the sample surface. a**, A schematic illustrating that the pump beam enters from the vacuum to the sample. **b**, A schematic describing the area illuminated by the pump beam on the sample surface when the sample is tilted.

## SE5. Analysis on momentum distribution curves at the X point

Analysis on momentum distribution curves (MDCs) was performed at the X point, as the similar analysis was done at the branch in the main text (Fig. 3). In Fig. SE5b and c, the extracted FWHM is overlaid on the E-k cuts. At PPm, the excited states cross $E_F$ and occupy the gap region. The spectral feature at PPm is qualitatively well captured by the calculated spectra from OC-DMFT simulations of $Sr_2IrO_4$ with the temperature and doping level of 800 K and 8% electrons/Ir-site, respectively, as shown in Fig. SE5e. Note that the FWHM profile at PPm is different from that at PPm – 3 ps (also, see Fig. SE7d). At PPm, as energy increases, the FWHM decreases and becomes the narrowest at ~ -0.10 eV and widens. However, at PPm – 3 ps, as energy increases, the FWHM monotonically decreases to ~ -0.03 eV. This indicates that the spectra at PPm are not from the surface photo-voltage effect of those at PPm – 3 ps, but from a genuine light-induced change in the electronic band structure.

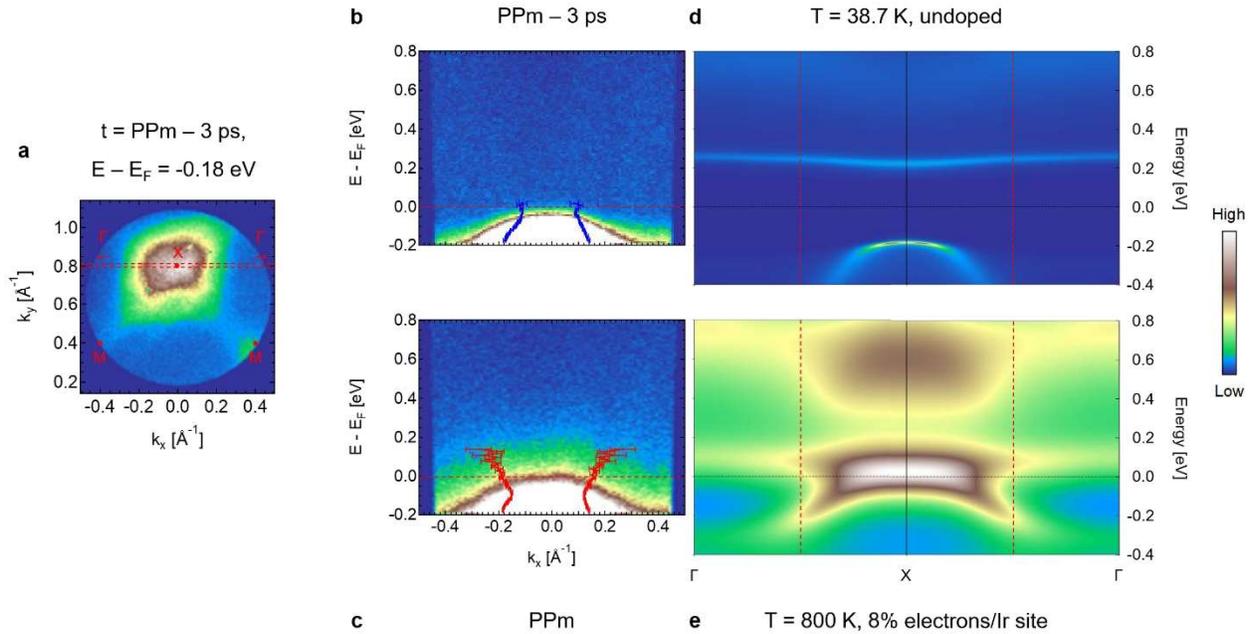

**Fig. SE5. Features of the electronic band structure at PPm – 3 ps and PPm at the X point. a**, A constant energy cut at PPm – 3 ps. The high symmetry points (the X and M points) and directions towards the Γ points are marked. **b, c**, E-k cuts at PPm – 3 ps (**b**) and PPm (**c**). The slicing location and the integration range to generate E-k cuts are indicated by two red dotted lines in **a**. Note that the color scale of two experimental E-k cuts is saturated to clearly show the weak signals at high energies, but the range of the color scale is the same for both E-k cuts. The overlaid lines show the FWHM (mean $\pm$ FWHM/2) of Gaussian fits on MDCs at each energy. Through these lines, the detailed shape of the electronic band structure can be captured. At PPm, the excited states cross $E_F$ and occupy the gap region. The electronic band structure at PPm (**c**) is clearly changed compared to that at PPm – 3 ps (**b**). **d, e**, Calculated E-k cuts at 38.7 K without doping (**d**) and 800 K with doping of 8% electrons / Ir-site (**e**). The slicing location is along Γ – X – Γ as in **b** and **c** (but no momentum integration). The two red dotted lines in **d** and **e** indicate the momentum range corresponding to the signal window in **b** and **c**. In **e**, the gap is closed and the electronic band structure crosses $E_F$, which reproduces the features at PPm in **c**. Note that the ranges of the color scales in **a, d**, and **e** are set from the minimum to the maximum values of the intensity of each panel. The color scales in all the panels (**a – e**) represent a linear scale.

## SE6. Extracted parameters from the MDC analysis at the branch and the X point

We performed Gaussian fittings on the momentum distribution curves (MDCs) at each energy in energy-momentum (E-k) cuts in the static spectrum and those from the delay time snapshots. It is challenging to visualize the shape of the excited states in E-k cuts since the signal of the excited states is inherently weak and the assigned color range for the weak signals is narrow. Although we can partially overcome this issue by saturating the color scale, as it is done in Figs. 3b and c, the extracted parameters from the Gaussian fittings provide a way to clearly visualize the detailed structure of the excited states and they show clear temporal dynamics. For example, by overlaying the FWHM (mean $\pm$ FWHM/2) from the Gaussian fittings, we could reveal the pillar shape of the excited states at PPm in Fig. 3c, and we could also observe a clear change of the band structure from PPm – 3 ps to PPm in Figs. 3b and c and Figs. SE5b and c. In panels d, e, and f in Figs. SE5 and SE6, we display all the parameters extracted from the Gaussian fittings on the MDCs at each energy in E-k cuts in the static spectrum and those from the delay time snapshots. In panels d and e in Figs. SE6 and SE7, the FWHM and the intensity show a clear temporal dynamics from the state before the system is pumped at PPm – 3 ps to the excited state at PPm and then to the relaxed state at PPm + 1 ps. Note that we extracted the electronic temperatures from the intensity-energy plots at PPm and PPm + 1 ps in Fig. SE7e with the method that is used in SE3. At PPm, 395 $\pm$ 10 K and 703 $\pm$ 76 K were extracted from the lower band (fitting range: E - $E_F$ = -0.172 eV ~ -0.052 eV) and the middle band (fitting range: E - $E_F$ = -0.052 eV ~ 0.048 eV), respectively. At PPm + 1 ps, 402 $\pm$ 5 K is extracted for the fitting range (in E - $E_F$) of -0.172 eV ~ 0.098 eV. All the values of the extracted electronic temperatures are close to those in SE3. This indicates that the momentum integration range for the EDCs used to extract the electronic temperatures in SE3 well includes the intensity peaks along the $k_x$ direction in the spectrum orientation shown in Fig. SE7a which has the same orientation with the spectrum in Fig. SE2b.

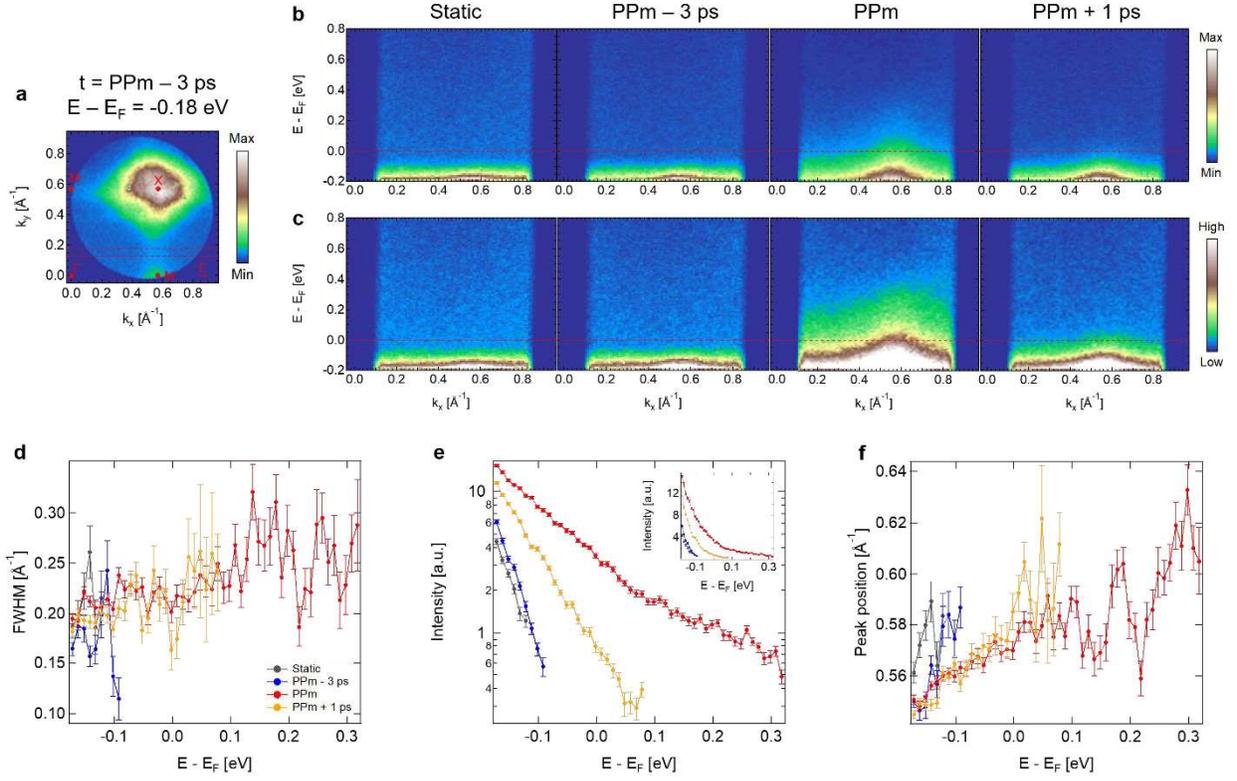

**Fig. SE6. Parameters extracted from the Gaussian fittings on MDCs at each energy in E-k cuts at the branch.**
**a**, A constant energy cut at PPm − 3 ps. The high symmetry points (the Γ, X, and M points) and the direction towards the Γ point are marked. The range of the color scale is set from the minimum to the maximum values of the intensity. **b**, **c**, E-k cuts for the static spectrum and those for delay time snapshots. The slicing location and the integration range to generate E-k cuts are indicated by two red dotted lines in **a**. E-k cuts in **b** and **c** are identical but have different color scales. In **b**, the ranges of the color scale are set from the minimum to the maximum values of the intensity of each panel. In **c**, the color scale of all E-k cuts is saturated to clearly show the weak signals at high energies and the range of the color scale is the same for all E-k cuts. The color scales in **a**, **b**, and **c** represent a linear scale. **d**, **e**, **f**, the extracted parameters (FWHM (**d**), intensity (**e**), and peak position (**f**)) from the Gaussian fittings on MDCs at each energy in E-k cuts. In **e**, curves are normalized by the total electron counts of each case. Since the total electron counts are on the order of $10^7$, we divided the total electron counts by $10^7$ and used these values for the normalization of each intensity plot. Note that the main panel is a semi-log plot, while the linear-scale version is shown in the inset. In **b** − **f**, there are clear temporal dynamics: at PPm, the excited states cross $E_F$ and there is a clear change in the band structure at PPm compared to that at PPm − 3 ps. At PPm + 1 ps, the excited states relax back and the gap is recovering.

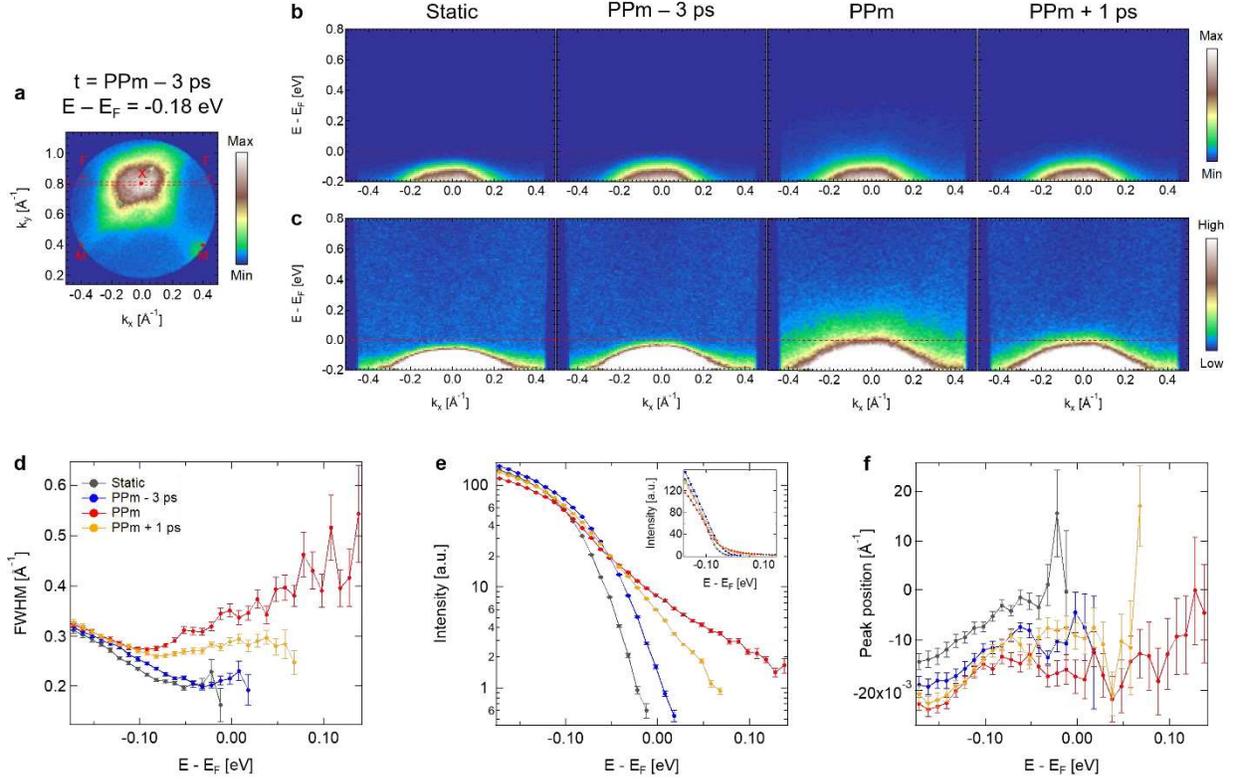

**Fig. SE7. Parameters extracted from the Gaussian fittings on MDCs at each energy in E-k cuts at the X point.**
**a**, A constant energy cut at PPm − 3 ps. The high symmetry points (the X and M points) and directions towards the Γ points are marked. The range of the color scale is set from the minimum to the maximum values of the intensity. **b**, **c**, E-k cuts for the static spectrum and those for delay time snapshots. The slicing location and the integration range to generate the E-k cuts are indicated by two red dotted lines in **a**. E-k cuts in **b** and **c** are identical but have different color scales. In **b**, the ranges of the color scale are set from the minimum to the maximum values of the intensity of each panel. In **c**, the color scale of all E-k cuts is saturated to clearly show the weak signals at high energies and the range of the color scale is the same for all E-k cuts. The color scales in **a**, **b**, and **c** represent a linear scale. **d**, **e**, **f**, the extracted parameters (FWHM (**d**), intensity (**e**), and peak position (**f**)) from the Gaussian fittings on MDCs at each energy in E-k cuts. In **e**, curves are normalized by the total electron counts of each case. Since the total electron counts are on the order of $10^7$, we divided the total electron counts by $10^7$ and use these values for the normalization of each intensity plot. Note that the main panel is a semi-log plot, while the linear-scale version is shown in the inset. In **b** − **f**, there are clear temporal dynamics: at PPm, the excited states cross $E_F$ and there is a clear change in the band structure at PPm compared to that at PPm − 3 ps, especially FWHM becomes the minimum at around -0.1 eV at PPm, while that at PPm − 3 ps shows a monotonically decreasing trend up to ∼ -0.03 eV. At PPm + 1 ps, the excited states relax back and the gap is recovering.

## SE7. Estimation of the Fermi level ($E_F$)

### SE7.1. Calibration of the Fermi level ($E_F$) with $Bi_2Se_3$ single crystals

We calibrated the Fermi level ($E_F$) by using metallic $Bi_2Se_3$ single crystals at 39.3 K. E-$k_x$ (Fig. SE8a) and E-$k_y$ cuts are generated at the Dirac point (the Γ point) and EDCs are acquired by using a narrow and a wide momentum range, as indicated by the pairs of the red and green dashed lines, respectively. For the fitting of the Fermi-Dirac function on EDCs, we used the following model.

$$I(E) = [(\text{DoS} \times f(T)) \otimes G](E) \qquad (24)$$

where I, DoS, f, $\otimes$, and G represent an EDC, density of states, the Fermi-Dirac distribution, the convolution operation, and a Gaussian function, respectively. First(linear)- and the second-order polynomials were used for the density of states. Fig. SE8b shows a fitting with the linear density of states. Red dotted lines indicate the fitting range and this range is also used for the other fittings. In total, there are eight different fits (E-$k_x$ and E-$k_y$ cuts, the narrow and the wide momentum integration ranges, and the first- and second-order polynomials for the density of states.). The average values for $E_F$ and the energy resolution are -0.033 eV (min = -0.035 eV and max = -0.031 eV) and 0.036 eV (min = 0.033 eV and max = 0.038 eV), respectively.

### SE7.2. Two data acquisition configurations: high energy acquisition and low energy acquisition

In the data acquisition, there are two important tuning knobs: the center energy ($E_c$) of the energy window (the range of the energy window is $\pm 3\%$ of $E_c$) and the high pass filter of the electron kinetic energy (DBIAS) which lets pass the electrons with high kinetic energy and filters out the electrons with low kinetic energy. By setting $E_c$ to a high value and suppressing most of the electrons from the valence band (high DBIAS), we can focus on the top part of the valence band: high energy configuration (high E config.). Considering the lifetime of the detector, we prefer limiting the electron count rate. In this condition, the high E config. has the advantage to maximize the signal from the excited states since most of the signal from the valence band is filtered out and the portion of the electron counts from the excited states becomes larger in the electron count rate. But the space charge effect can build up since the total electron count rate before the high pass filter is large. The data in Figs. 2 and 3 were taken with the high E config. Another configuration is setting $E_c$ to a low value and lower the energy cut for filtering the electron kinetic energy (low DBIAS): low energy configuration (low E config.). In this case, we can see the structure of the valence band down to deeper energies. But the electron count rate is mostly dominated by the valence band electrons, so the signal from the excited states is very weak. The data in Fig. 1 were taken with the low E config. We found that there is a slight energy shift of the leading edges between the EDCs from the high E config. and the low E config., as shown in Fig. SE8d. With proper scaling and translation factors, we could match the leading edges of the EDCs as shown in Figs. SE8e and f, and the amount of the energy shift is 0.030 eV. This energy shift value is consistent in the spectra from the two configurations for static, PPm − 3 ps, PPm, and PPm + 1 ps. Note that, in Figs. SE8c − f, a hump feature between -0.1 eV and -0.2 eV in the EDC from the low E config. does not exist in that from the high E config., which is possibly due to a worsening of the energy resolution in the high E config. by the space charge effect. The reason why there is an energy shift between spectra from the two configurations is currently not clear. Since the $Bi_2Se_3$ spectra were taken with the low E config., we directly apply the calibrated $E_F$ value from the $Bi_2Se_3$ spectra to the $Sr_2IrO_4$ spectra taken with the low E config. (e.g. Fig. 1). But for the $E_F$ value in the $Sr_2IrO_4$ spectra taken with the high config. (e.g. Fig. 2 and 3), we additionally consider a 0.030 eV shift on top of the calibrated $E_F$ value from the $Bi_2Se_3$ spectra. Note that the detailed numbers in $E_c$ and DBIAS in the low E config. for the $Bi_2Se_3$ spectra and the $Sr_2IrO_4$ spectra are different, but the energy cut for filtering the electron kinetic energy is far from $E_F$ in both cases.

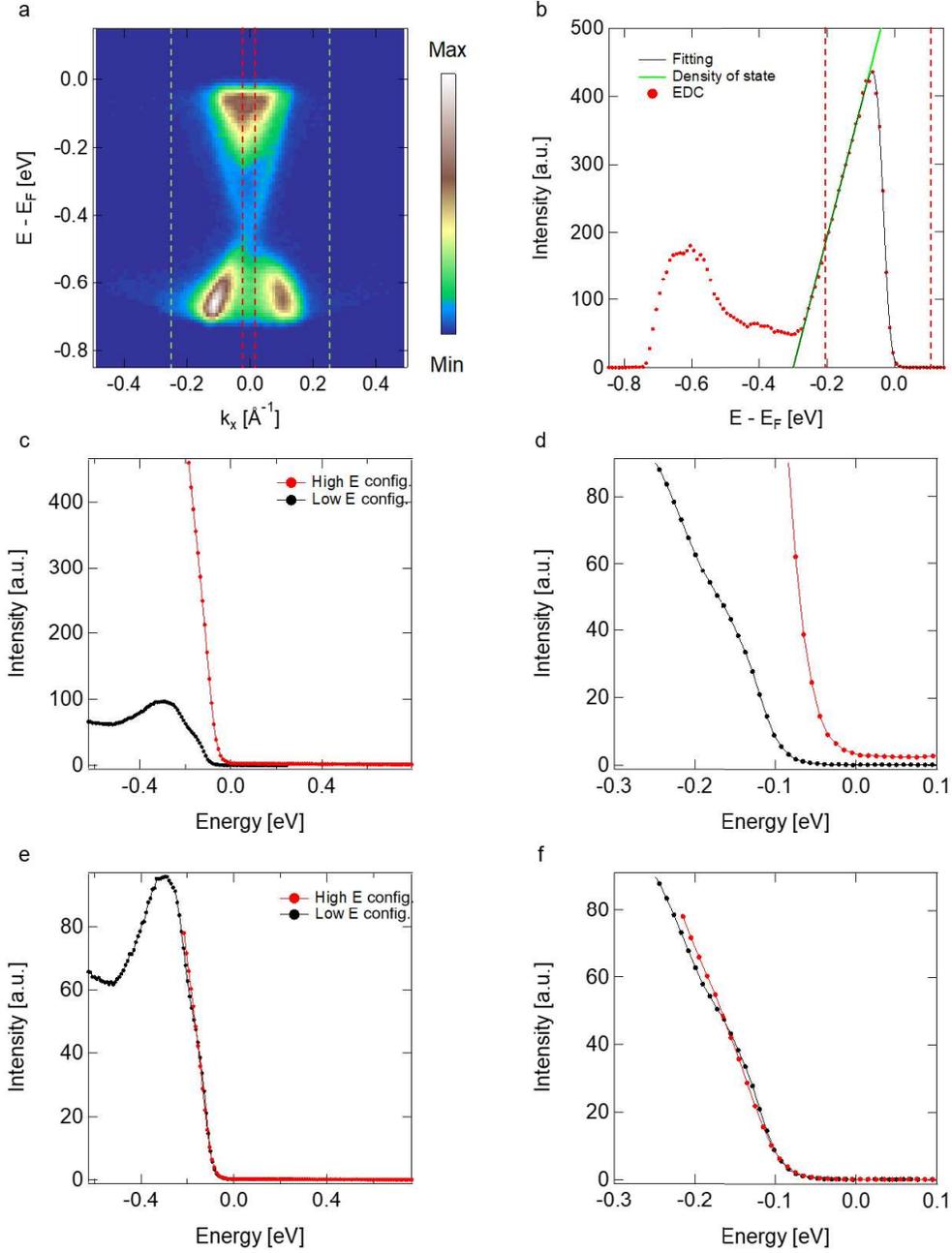

**Fig. SE8. Calibration of the Fermi level ($E_F$) a**, E-$k_x$ cut at the Dirac point (the $\Gamma$ point) in the $Bi_2Se_3$ spectra. **b**, EDC generated by integrating the momentum range indicated by two red dotted lines in **a**. A linear function is used for the density of states. Fitting was performed using the model in Eq. (24). The fitting range is indicated by two red dotted lines. We performed the fitting on EDCs from E-$k_x$ and E-$k_y$ cuts with the narrow and the wide momentum integration ranges (indicated by the two red and green dashed lines in **a**, respectively.). First(linear)- and second-order polynomials were used for the density of states. In total, there are eight fits (E-$k_x$ and E-$k_y$ cuts, the narrow and the wide momentum integration ranges, and the first- and second-order polynomials for the density of states.). **c**, EDCs at the X point from the static spectra with the high E config. and the low E config. (red and black colors, respectively). The momentum integration range to generate the EDCs is shown in Fig. SE2b. $E_F$ is not calibrated. **d**, The ranges of the X and Y axes in **c** are zoomed in to see the leading edges. **e**, EDC from the high E config. is scaled and translated with the factors of Y multiplication = 0.17, X offset = -0.03 eV, Y offset = -0.2. **f**, The ranges of the X and Y axis in **e** are zoomed in to see the leading edges. After the scaling and the translation, the leading edges from the two EDCs

match. A hump feature between -0.1 eV and -0.2 eV in the EDC from the low E config. does not exist in that from the high E config., which is possibly due to the worsening of energy resolution in the high E config. by the space charge effect.

## SE8. Delay time scan and the delay time for the maximum intensity of the excited states (PPm)

To determine the delay time for the maximum intensity of the excited states (pump–probe signal maximum, PPm), we acquired spectra as we sweep the delay time. A clear signal of the excited states near the pump–probe temporal overlap (0 ps) is shown in Fig. SE9a. The delay time for the pump–probe temporal overlap was determined by fitting a Gaussian function in the intensity–delay time cut at 0.7 eV in Fig. SE9b. The photon energy of the pump beam is 1.2 eV, but the upper limit of the energy window is ~ 0.8 eV. So, we decided to get the cut (Fig. SE9b) at 0.7 eV where the excitation and decay are still quick and well-fitted with the Gaussian function, which can be used to determine the delay time for the pump–probe temporal overlap. The FWHM of the Gaussian fitting is about 0.40 ps, which can be considered as an upper bound of the time resolution of our measurement. We decided PPm to be at 0.14 ps based on the intensity–delay time cut at 0 eV, shown in Fig. SE9c.

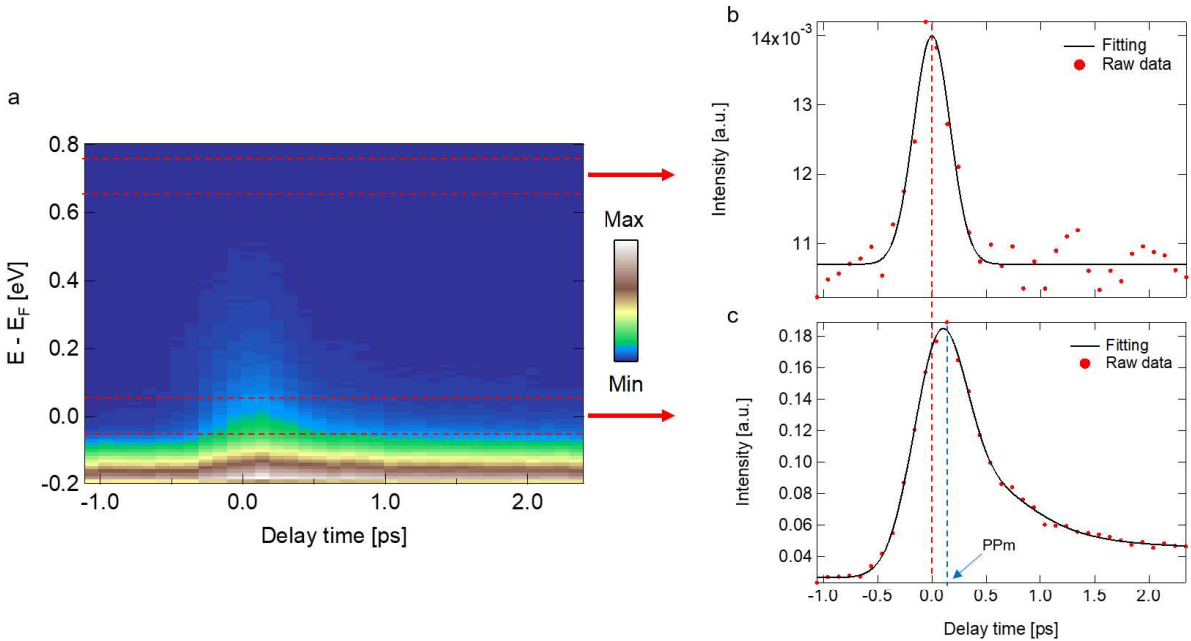

**Fig. SE9. Delay time scan and the delay time for the maximum intensity of the excited states (PPm). a**, Energy – delay time plot. The whole momentum range in the measurement window was integrated. The range of the color scale is set from the minimum to the maximum values of the intensity. **b**, Intensity – delay time plot at 0.7 eV. The energy position and the integration range are shown by the upper two red dotted lines in **a**. A Gaussian function is used to fit the raw data and the peak position of the Gaussian fitting is used for the delay time for the pump–probe temporal overlap. **c**, Intensity – delay time plot at 0 eV. The energy position and the integration range are shown by the lower two red dotted lines in **a**. The delay time for the maximum intensity of the excited states (pump-probe signal maximum, PPm) is at 0.14 ps and marked by the blue dotted line. The fitting is for the guidance of the eyes and we used an error function and a bi-exponential function (= [$A_1$ + 0.5 × (1 + erf function) × ($A_2$ + two exponential functions)] where $A_1$ and $A_2$ are constants) as the fitting function.

**SE9. Advantages of trARPES to explore the states above $E_F$ and at high temperatures.**

There are two main advantages of trARPES[19] for the purpose of this study. First, trARPES gives access to states above $E_F$ and enables us to directly observe whether the gap is closed as a result of high electronic temperature and/or photo-doping. Our data provide unambiguous evidence demonstrating the appearance of bands crossing $E_F$ and extending beyond the energy where the conduction band is expected to be located. Next, trARPES allows us to examine the effect of very high electronic temperatures (and photo-doping) on the electronic structure of $Sr_2IrO_4$ with cold lattice temperature. In our study, to evaluate the nature of the gap, it is imperative to access the electronic band structure at high electronic temperature (with photo-doping). But also, in the broader context of insulator-metal transitions (IMTs), the ability to access a high (electronic) temperature is crucial since the temperature is an important parameter in the IMT process. In addition, the temperature effect can be easily studied in the model calculations. Since the electronic specific heat is usually very small, when a pump pulse is applied, the electronic temperature almost instantaneously reaches very high values -- here we estimate 400 K - 2100 K at PPm (see SE3) -- and it takes several hundreds of femto-second for a significant amount of energy to be transferred from the electrons to the lattice through electron-phonon scattering. Through pump pulses with a short enough temporal width (ours is ~135 fs) compared to the time needed for the energy in the electrons to be transferred to the lattice, the effect of the high electronic temperature with a cold lattice can be explored. In addition, in our trARPES experiments, it is inevitable to have both a high electronic temperature and photo-doping effects, since we pump electrons above the gap with the 1.2 eV pump beam. On the other hand, in static ARPES, the system is measured in an equilibrium state, where the electronic and lattice temperatures are necessarily the same. Technically, with static ARPES, it is challenging to investigate the temperature dependence of the electronic band structure while heating the sample to a very high temperature, e.g. 1000 K, since increasing temperature to such a high value significantly worsens the vacuum level and deteriorates the quality of the cleaved surface of the sample. However, applying pump pulses provides local heating in the sample so that it does not (significantly) degrade the vacuum level. The ability to measure states above $E_F$ with high electronic temperature and/or photo-doping distinguishes our data from the results of low-temperature static measurements on chemically doped $Sr_2IrO_4$.

## SE10. Surface photovoltage effect at PPm − 3 ps

When a probe pulse arrives on the sample surface before a pump pulse (negative delay time), an electron is photoemitted and flies toward the detector in the vacuum. After that, when the pump pulse arrives on the sample surface, electron-hole pairs are generated. Those electrons and holes are spatially separated due to the electric field by the surface band bending, followed by the generation of the long-range dipole field which accelerates (or decelerates) the electron flying in the vacuum. This effect is called the surface-photovoltage effect (SPV)[20] and it is manifested by the shift of the band in the energy axis. In Fig. SE10, we compared two EDCs at the X point from the static spectrum and the one at PPm − 3 ps. With proper scaling and translation, the overall feature of the two EDCs well overlaps as shown in Fig. SE10b. This means SPV shifts up the spectra at PPm − 3 ps by ~ 10 meV. Note that there is a mismatch in two EDCs at around -0.15 eV. In pump-probe experiments, there is a steady-state heating by the pump pulses, and this can result in the thermal broadening of the band. The faint spectral feature in Fig. 2a at $E_F$ at PPm − 3 ps is attributed to the SPV and thermal broadening of the valence band due to the steady-state heating by the pump pulses.

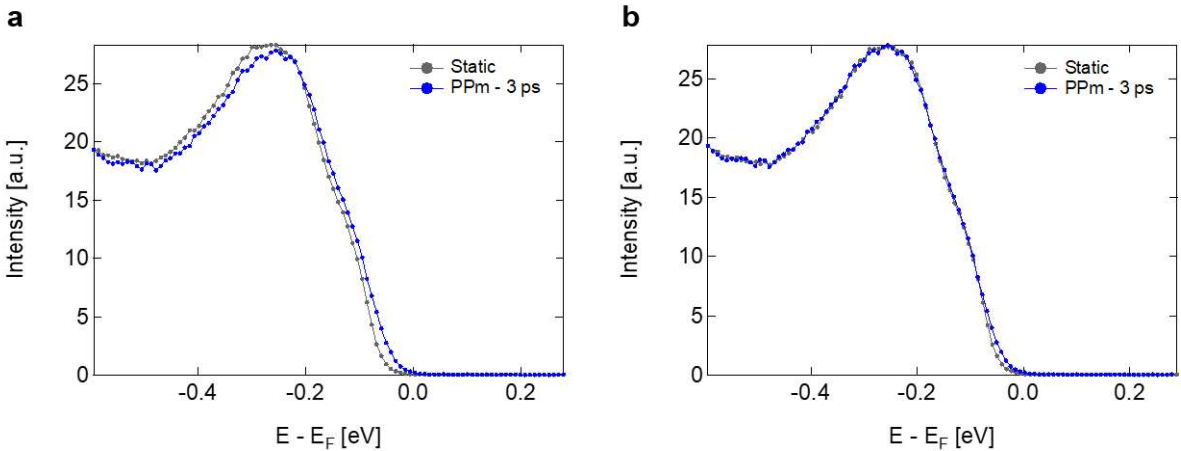

**Fig. SE10. a**, EDCs at the X point from the static spectrum and the one at PPm − 3 ps. Note that the spectra were acquired with the low E config. (see SE6.2). The momentum integration area for the EDCs at the X point is shown in Fig. SE2b. The EDCs are normalized by the total electron counts of each spectrum. Since the total electron counts are on the order of $10^7$, we divided the total electron counts by $10^7$ and use these values for the normalization of each EDC. **b**, Scaling and translation are applied on the EDC from the static spectrum (X offset = 0.01 eV, Y multiplication = 0.978). The overall features of two EDCs well overlap and this means that the SPV shifts up the spectra at PPm − 3 ps by ~ 10 meV. Note that there is a mismatch in the two EDCs at around -0.05 eV. In pump-probe experiments, there is a steady-state heating by the pump pulses, and this can result in the thermal broadening of the band.

**SE11. Low sensitivity area near the center of the measurement window**

There is an area with low sensitivity near the center of the measurement window. It is indicated by a red arrow in Fig. SE11. We avoided this low sensitivity area when we integrate the momentum ranges to generate E − k cuts, EDCs, and MDCs.

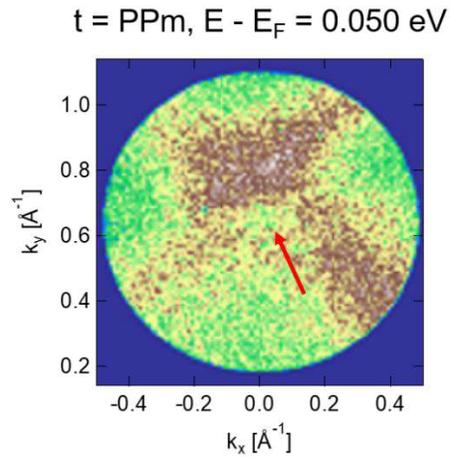

**Fig. SE11.** A constant energy cut at E − $E_F$ = 0.050 eV at PPm. The area with low sensitivity is indicated by the red arrow.

## SE12. Analysis of the spectral intensity in the gap region at PPm: gap-closing

For the evolution of the band gap in a perturbed system (e.g. by charge doping or photo-excitation), one commonly distinguishes the scenarios of gap-closing and gap-filling (which are well depicted in Fig. 2a and b in ref. [21]). Gap-closing refers to a reduction of the distance between two bands (e.g. conduction and valence bands) which are separated by the gap. This effect can be tracked by the evolution of the peak positions of the two bands. In contrast, gap-filling refers to a broadening of the two bands, in particular due to thermal effects, while the gap size and the peak positions of the two bands are preserved.

In Fig. SE12b, the EDCs at the X point are compared for the spectrum at PPm − 3 ps and that at PPm. In the EDC at PPm, the peak at -0.24 eV slightly moves to more negative energies, while the hump at ∼ -0.10 eV does not move compared to the EDC from the spectrum at PPm − 3 ps. Note that the hump feature is clearer in the translated EDC from the static spectrum in Fig. SE12c. These behaviors are not consistent with an ideal gap-closing scenario, while the excited states of the high energy tail of the EDC at PPm may at first sight resemble a standard gap-filling. Note that it is not clear which peak corresponds to the $J_{eff}$ = ½ states (see SE1), so one of the two features may not be relevant to the current discussion. However, the behavior of the excited states in the gap region at PPm differs from a thermal gap-filling scenario in several important respects: (1) First, the energy vs. momentum dispersion of the excited states at the branch and the X point, as disclosed by the FWHM from the MDC analysis, has a non-trivial form. In particular, we would like to highlight four important features: (1-1) The biconcave shape of the dispersion of the excited states at the X point (Fig. SE5c) cannot be obtained from the one at PPm − 3 ps (Fig. SE5b) by thermal broadening of the valence band. If the dispersion of the excited states would be derived from a thermal broadening of the static band, the shape of the dispersion should remain the same as the one at PPm − 3 ps. (1-2) The pillar shape at the branch (Fig. 3c) and the biconcave shape at the X point (Fig. SE5c) of the excited states are qualitatively reproduced as electronic bands (not a thermal broadening of the other bands) by the simulation (Fig. 3e and Fig. SE5e, respectively) and tight-binding calculation with U = 0 eV[2]. (1-3) Also, the pillar shape of the excited states at the branch (Fig. 3c) resembles the metallic band at the M point after the gap is collapsed in 5% La-doped $Sr_2IrO_4$ at low temperature (thermal broadening is negligible), measured via static ARPES[2]. (1-4) It is also important to note that the shape of the dispersion of the excited states at the branch and the X point evolves in different ways (pillar-like vs. biconcave). If those were the result of a thermal broadening of the static bands, then they should evolve in a similar fashion. (2) Next, there is no signature of the conduction band onset in the EDCs at the branch and the X point at PPm in Fig. 2b and c. The shape of the excited states in the EDCs at PPm rather resembles the non-thermal electron profile within the metallic density of states. For the above reasons, we conclude that the excited states appearing in the gap region at PPm cannot originate from a trivial gap-filling via thermal broadening of the bands.

Although the electronic spectrum below $E_F$ does not look consistent with the gap-closing scenario after photo-excitation (persistent hump at ∼ -0.10 eV and slight shift of the peak at -0.24 eV to more negative energies), the above-mentioned observations (1-2), (1-3), and (2) on the excited states as a whole clearly support the gap-closing scenario. We thus interpret the excited states appearing in the gap region at PPm as originating from a gap-closing or transient insulator-metal transition induced by light excitation. This in particular explains the non-trivial evolution of the dispersion of the excited states, since the transient metallic state at PPm can have a density of states in the gap region associated with a characteristic band dispersion.

Further investigations are needed to understand how the behavior of the hump at ∼ -0.10 eV and the peak at -0.24 eV at the X point at PPm can be reconciled with the gap-closing picture. There are different possible interpretations: (1) The spectra at PPm may be a superposition of spectra from metallic and insulating regions associated with a spatially inhomogeneous evolution from the gapped to the gapless state. Spatial inhomogeneity with gapped and gapless regions has been revealed by STM studies on La-doped $Sr_2IrO_4$[22,23], while the theoretical calculations and simulations in our study are for a homogeneous system. However, static ARPES on La-doped $Sr_2IrO_4$ with a similar doping level yields a metallic spectrum without indications for the superposition of these spectra with those from insulating states[2]. (2) Since the electronic system is not thermalized at PPm, it may exhibit unique non-equilibrium features which do not exist in thermal spectra. For example, the band below $E_F$ may survive the pump excitation, while the conduction band collapses, generating new states in the gap region. Such a nonequilibrium state could be approximated by a combination of the valence band of the insulating state and the bands from the metallic states generated via electron doping and high temperature. Future simulations of the electronic structure with non-thermal electronic distributions may shed more light on this relevant question.

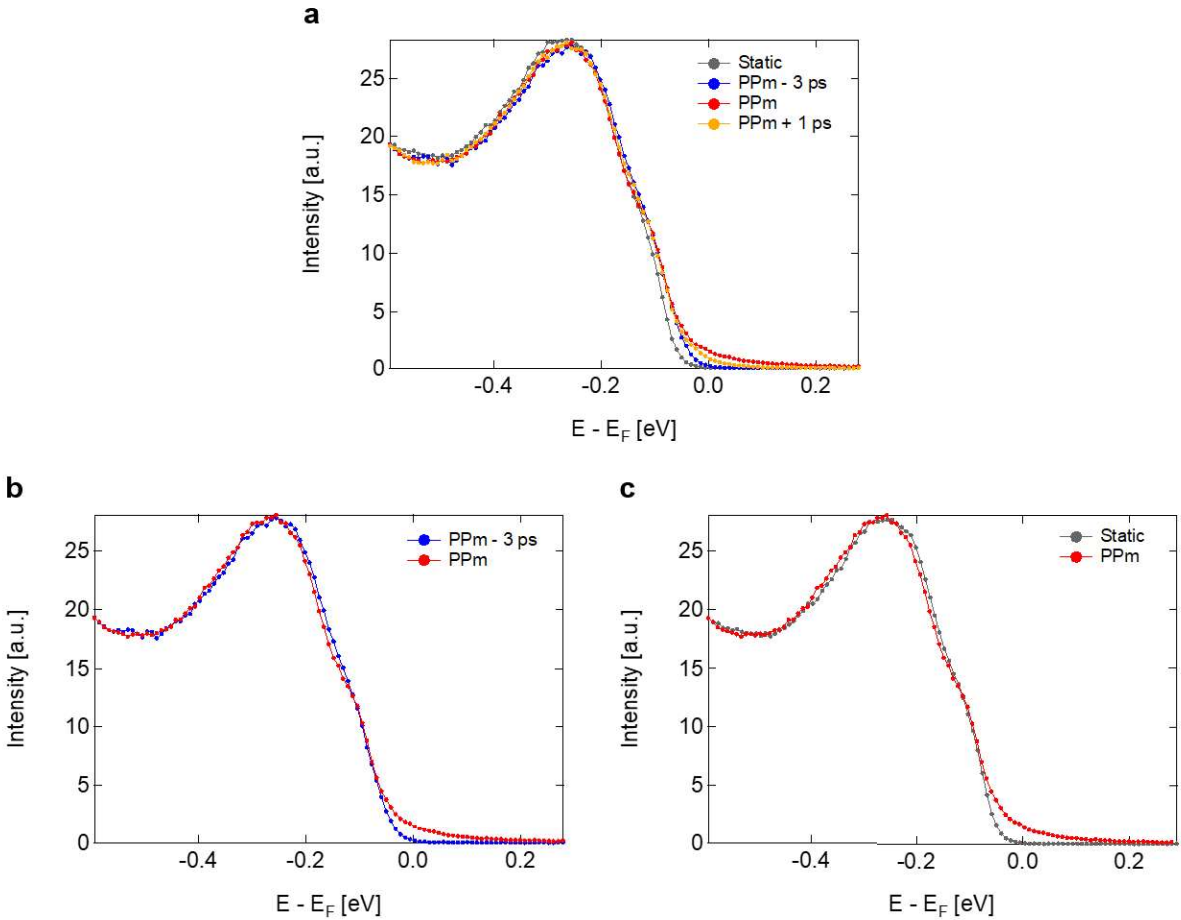

**Fig. SE12. a**, EDCs at the X point from the static spectrum and those at different delay times. Note that the spectra were acquired with the low E config. (see SE6.2). The momentum integration area for the EDCs at the X point is shown in Fig. SE2b. The EDCs are normalized by the total electron counts of each spectrum. Since the total electron counts are on the order of $10^7$, we divided the total electron counts by $10^7$ and use these values for the normalization of each EDC. **b**, **c**, the EDC at PPm is compared to that at PPm − 3 ps (**b**) and that from the static spectrum (**c**). Note that the EDC from the static spectrum in **c** is the same as the one in Fig. SE10b: it is translated in energy by 10 meV and its intensity is scaled to match to that at PPm − 3 ps. Since the EDC from PPm − 3 ps shows a less clear hump (at ∼ -0.10 eV, see Fig. SE10a) and broadened tail near $E_F$, to clearly observe the change of features at PPm, we need to compare the EDC at PPm to the translated EDC from the static spectrum. In **b** and **c**, the peak position of the large peak at -0.24 eV at PPm shifts slightly down in energy. The hump at ~ -0.10 eV becomes clearer at PPm. But the hump position is not changed. There is clear spectral weight above $E_F$ at PPm. For the discussion of gap-closing vs. gap-filling, see SE11.